\documentclass[twocolumn]{aastex631}
\graphicspath{{./fig/}{./png/}}
\usepackage{bm}

\newcommand{\EQ}{\begin{equation}}
\newcommand{\EN}{\end{equation}}
\newcommand{\EQA}{\begin{eqnarray}}
\newcommand{\ENA}{\end{eqnarray}}

\newcommand{\Eq}[1]{Equation~(\ref{#1})}
\newcommand{\Eqs}[2]{Equations~(\ref{#1}) and~(\ref{#2})}

\newcommand{\Sec}[1]{Section~\ref{#1}}

\newcommand{\Fig}[1]{Figure~\ref{#1}}

\newcommand{\Figs}[2]{Figures~\ref{#1} and \ref{#2}}

\newcommand{\Tab}[1]{Table~\ref{#1}}

\newcommand{\bra}[1]{\langle #1\rangle}
\newcommand{\bbra}[1]{\left\langle #1\right\rangle}

\newcommand{\meanPhi}{\overline{\Phi}}
{}
{}
\newcommand{\meanFFFF}{\overline{\mbox{\boldmath ${\cal F}$}}{}}{}

{}
{}
\newcommand{\meanEMF}{\overline{\mbox{\boldmath ${\cal E}$}}{}}{}
{}
{}
{}
{}
{}
\newcommand{\meanEE}{\overline{\mbox{\boldmath $E$}}{}}{}
{}
{}
{}
{}
{}
{}
{}
{}
\newcommand{\meanUU}{\overline{\bm{U}}}

{}

\newcommand{\meanA}{\overline{A}}
\newcommand{\meanB}{\overline{B}}

\newcommand{\meanU}{\overline{U}}

\newcommand{\meanFFFmz}{\overline{\cal F}_{\mathrm{m}z}}
\newcommand{\meanFFFfz}{\overline{\cal F}_{\mathrm{f}z}}

{}

{}
{}
{}

\newcommand{\xxx}{\hat{\mbox{\boldmath $x$}} {}}
\newcommand{\yyy}{\hat{\mbox{\boldmath $y$}} {}}

\newcommand{\meanAA}{{\overline{\bm{A}}}}
\newcommand{\meanBB}{{\overline{\bm{B}}}}
\newcommand{\meanJJ}{{\overline{\bm{J}}}}

\newcommand{\kk}{\bm{k}}

\newcommand{\xx}{\bm{x}}
\newcommand{\aaaa}{\bm{a}}
\newcommand{\jj}{\bm{j}}

\newcommand{\bb}{\bm{b}}
\newcommand{\BB}{\bm{B}}
\newcommand{\aaT}{\bm{a}_\mathrm{T}}
\newcommand{\bbT}{\bm{b}_\mathrm{T}}
\newcommand{\BBT}{\overline{\bm{B}}_\mathrm{T}}

\newcommand{\EE}{\bm{E}}
\newcommand{\JJ}{\bm{J}}
\newcommand{\oo}{\bm{\omega}}

\newcommand{\AAA}{\bm{A}}

\newcommand{\UU}{\bm{U}}
\newcommand{\VV}{\bm{V}}

\newcommand{\uu}{\bm{u}}

\newcommand{\ee}{\mbox{\boldmath $e$} {}}

\newcommand{\ff}{\mbox{\boldmath $f$} {}}

\newcommand{\nab}{{\bm{\nabla}}}

\newcommand{\OO}{\bm{\Omega}}

\newcommand{\ddelta}{\mbox{\boldmath $\delta$} {}}
\newcommand{\ggamma}{\mbox{\boldmath $\gamma$} {}}

\newcommand{\SSSS}{\mbox{\boldmath ${\sf S}$} {}}

\newcommand{\ii}{{\rm i}}

\newcommand{\DDD}{{\cal D} {}}

\newcommand{\dd}{{\rm d} {}}

\def\la{\mathrel{\mathchoice {\vcenter{\offinterlineskip\halign{\hfil
$\displaystyle##$\hfil\cr<\cr\sim\cr}}}
{\vcenter{\offinterlineskip\halign{\hfil$\textstyle##$\hfil\cr<\cr\sim\cr}}}
{\vcenter{\offinterlineskip\halign{\hfil$\scriptstyle##$\hfil\cr<\cr\sim\cr}}}
{\vcenter{\offinterlineskip\halign{\hfil$\scriptscriptstyle##$\hfil\cr<\cr\sim\cr}}}}}

\def\H{\mbox{\rm H}}

\def\Ma{\mbox{\rm Ma}}
\def\Co{\mbox{\rm Co}}

\def\Shear{\mbox{\rm Sh}}

\def\Pm{\mbox{\rm Pr}_{\rm M}}
\def\Rm{\mbox{\rm Re}_{\rm M}}

\def\Rey{\mbox{\rm Re}}

\def\Co{\mbox{\rm Co}}

\def\cs{c_{\rm s}}

\def\vA{v_{\rm A}}

\def\kT{k_{\rm T}}

\def\kf{k_{\rm f}}

\def\epsf{\epsilon_{\rm f}}
\def\epsfz{\epsilon_{\rm f0}}
\def\epsfzM{\epsilon_{\rm f0}^\mathrm{M}}

\def\meanBrms{\overline{B}_{\rm rms}}
\def\Brms{B_{\rm rms}}

\def\urms{u_{\rm rms}}

\def\kappat{\kappa_{\rm t}}

\def\etat{\eta_{\rm t}}
\def\etatz{\eta_{\rm t0}}
\def\etatz{\eta_{\rm t0}}

\def\etaT{\eta_{\rm T}}

\def\Beq{B_{\rm eq}}

\def\half{{\textstyle{1\over2}}}

\newcommand{\W}{\,{\rm W}}
\newcommand{\V}{\,{\rm V}}

\newcommand{\T}{\,{\rm T}}
\newcommand{\G}{\,{\rm G}}

\newcommand{\K}{\,{\rm K}}
\newcommand{\g}{\,{\rm g}}
\newcommand{\s}{\,{\rm s}}

\newcommand{\m}{\,{\rm m}}

\newcommand{\J}{\,{\rm J}}

\newcommand{\A}{\,{\rm A}}

\def\apjl{ApJL}
\def\apjs{ApJS}

\def\aap{A\&A}

\begin{document}

\title{Magnetic helicity fluxes in dynamos from rotating inhomogeneous turbulence}

\email{brandenb@nordita.org}
\author[0000-0002-7304-021X]{Axel Brandenburg}
\affiliation{Nordita, KTH Royal Institute of Technology and Stockholm University, Hannes Alfv\'ens v\"ag 12, SE-10691 Stockholm, Sweden}
\affiliation{The Oskar Klein Centre, Department of Astronomy, Stockholm University, AlbaNova, SE-10691 Stockholm, Sweden}
\affiliation{McWilliams Center for Cosmology \& Department of Physics, Carnegie Mellon University, Pittsburgh, PA 15213, USA}
\affiliation{School of Natural Sciences and Medicine, Ilia State University, 3-5 Cholokashvili Avenue, 0194 Tbilisi, Georgia}

\author[0000-0002-2307-3857]{Ethan T. Vishniac}
\affiliation{Physics and Astronomy Department, Johns Hopkins University, Baltimore, MD 21218, USA}

\begin{abstract}
We analyze direct numerical simulations of large-scale dynamos in
inhomogeneous nonhelically driven rotating turbulence with and without
shear.
The forcing is modulated so that the turbulent intensity peaks in the
middle of the computational domain and drops to nearly zero at the two
ends above and below the midplane.
A large-scale dynamo is driven by an $\alpha$ effect of
opposite signs in the two hemispheres.
In the presence of shear, the hemispheric magnetic helicity flux from
small-scale fields becomes important and can even overcompensate for the magnetic
helicity transferred by the $\alpha$ effect between large and small scales.
This effect has not previously been observed in nonshearing simulations.
Our numerical simulations show that the hemispheric magnetic helicity
fluxes are nearly independent of the magnetic Reynolds number, but those
between large and small scales, and the consequent dynamo effect, are
still found to decrease with increasing Reynolds number---just like in
nonshearing dynamos.
However, in contrast to nonshearing dynamos, where the generated mean
magnetic field declines with increasing magnetic Reynolds number, it is
now found to remain independent of it.
This suggests that catastrophic dynamo quenching is alleviated by the
shear-induced hemispheric small-scale magnetic helicity fluxes that
can even overcompensate the fluxes between large and small scales and
thereby cause resistive contributions.
\end{abstract}


\section{Introduction} \label{sec:intro}

Many astrophysical bodies harbor large-scale magnetic fields.
Late-type stars and galaxies are the main examples where a dynamo converts
kinetic energy into large-scale magnetic energy \citep{Charbonneau14, BN23}.
Disks around young stars and compact objects are additional examples,
where the existence of large-scale magnetic fields has so far only been
inferred from simulations \citep{Armitage11,Jiang+14,DT20}.
In all these cases there is turbulence, the bodies rotate, and they are
stratified in the sense that the density and/or the turbulent velocity
vary in space.
This, together with the overall rotation, causes the turbulence to
become helical, which leads to an $\alpha$ effect \citep{Par55,SKR66},
where the coefficient $\alpha$ relates the mean electromotive force to
the mean magnetic field.
Also, the magnetic field attains helicity, which affects the $\alpha$
effect \citep{PFL76}.
The underlying theory has been the subject of textbooks \citep{Mof78,
Par79, KR80, ZRS83}, but later it became clear that magnetic helicity
conservation causes such dynamos to saturate at progressively lower values
as the microphysical resistivity decreases or the conductivity increases
\citep{GD96, Ji99, Kleeorin+00, Bra01, Vishniac+Cho01, FB02, BB02}.
The resulting mean magnetic field would then be very weak for
astrophysically relevant resistivities.

\cite{BF00} coined the term catastrophic quenching, which in its
original form refers to the actual value of $\alpha$ becoming very
small at low resistivities.
In particular, numerical simulations by \cite{CH96} have shown that for
mean fields defined as volume averages, the value of $\alpha$ diminishes
to zero as the conductivity increases.
It was therefore thought to be difficult to explain the generation of
the large-scale magnetic fields observed in many astrophysical bodies
with an $\alpha$ effect.
It quickly became clear that the problem of catastrophic quenching
is connected with the homogeneity of the turbulence in such simple
numerical setups.
In those cases, there can be no magnetic helicity flux and magnetic
helicity is then well conserved in the limit of large conductivity.
To avoid this difficulty, \cite{Vishniac+Cho01} envisaged an $\alpha$
effect that is computed from a specifically designed magnetic helicity
flux such that the magnetic helicity is conserved exactly.
However, the anticipated magnetic helicity fluxes have not yet been
found in numerical simulations \citep{HB12}.
With just inhomogeneous turbulence, many numerical simulations show
that the amplitude of the resulting mean magnetic field decreases with
increasing conductivity \citep{DSGB13,Rin21,Bermudez+Alexakis22}.
This phenomenon is then generally also still referred to as the
catastrophic quenching problem, even though $\alpha$ itself may no longer
be catastrophically small.
On the other hand, corresponding analytic calculations of the mean
magnetic helicity fluxes by \cite{KR22} have not shown a dependence
of the magnetic helicity flux on the microphysical conductivity.
The reason for catastrophic quenching remains therefore obscure.

Many previous numerical simulations have employed helical forcing.
The purpose of the present work is to avoid this by adopting a more
natural setup in which a nonhelical flow is being driven.
Kinetic helicity of the flow can then emerge self-consistently as a
result of stratification and rotation.
We also consider the effect of shear and how it contributes to the
$\alpha$ effect \citep{RB14}.
Shear may also be responsible for driving magnetic helicity fluxes
\citep{Vishniac+Cho01}.

In \Sec{sec:model}, we describe our numerical simulations and the
test-field method that is used to compute the turbulent transfer coefficient.
We also discuss the decomposition of magnetic helicity fluxes into
contributions between hemispheres and between large and small scales.
In \Sec{sec:Results}, we describe the results without and with shear.
We conclude in \Sec{sec:Concl}.

\section{Description of the model}
\label{sec:model}

\subsection{Setup of our model} \label{sec:setup}

In this paper, we focus on the analysis of the magnetic helicity fluxes
resulting from a simulation in slab geometry with horizontal $xy$
averages depending on time and disk height $z$.
In the middle of the domain at $z=0$, the averaged turbulent intensity
has a maximum.
The angular velocity vector points in the positive $z$-direction, which
allows us to associate the regions above and below the midplane with
north and south.
This geometry can also be applied to the two sides around the equator
of a sphere, where the $z$-coordinate corresponds to latitude.

For our numerical simulations, we employ the \textsc{Pencil Code} \citep{PC}.
Since it advances the magnetic vector potential, the magnetic field is
always solenoid and the code is well suited for the task of analyzing
magnetic helicity and its flux.
Other codes that instead evolve the magnetic field and use divergence
cleaning to keep the magnetic field solenoidal can spontaneously produce
or destroy small-scale magnetic helicity \citep{BS20}, although schemes
have been devised to conserve magnetic helicity at the expense of
modifying the magnetic field in neighboring places \citep{ZV23}.
The \textsc{Pencil Code} has also been used successfully in various
earlier studies of magnetic helicity fluxes \citep{HB10, HB11, DSGB13,
Bra18}.

\subsection{Governing equations} \label{sec:equations}

We consider nonhelically driven inhomogeneous turbulence of an isothermal
gas with constant sound speed $\cs$ in the presence of rotation with the
angular velocity vector $\OO\equiv(0,0,\Omega)$.
In some cases, we include an additional shear flow, $\VV=(0,Sx,0)$,
where $S=-q\Omega$ is a constant and $q$ is the shear parameter.
Shear flows with $q<2$ are Rayleigh stable, but unstable to the
magnetorotational instability for $q>0$; see \cite{BH98} for a review.
Keplerian shear corresponds to $q=3/2$, while shear in galactic disks
corresponds to $q=1$ \citep{Beck+96}.
The turbulence is stochastically driven with a forcing function
$\ff(\xx,t)$, whose intensity is modulated in the $z$-direction
with a profile function $f_\mathrm{prof}(z)=\half(1+\cos k_1 z)$, where
$k_1=2\pi/L$ is the lowest wavenumber in a cube of size $L^3$.
To assess the sensitivity of the results upon this
choice, we also consider a case with a top-hat profile by using
$f_\mathrm{prof}(z)=\half[1+\tanh(5\cos k_1 z)]$, which has steep flanks
at $z=\pm\pi$, as quantified by the factor $5$ in front of the
cosine function.
The forcing is applied on the right-hand side of the evolution equation
for the velocity $\UU$, which then reads \citep{BNST95, BRRK08}
\begin{eqnarray}
\frac{\DDD\UU}{\DDD t}=&&\ff(\xx,t)-\UU\cdot\nab\UU-SU_x\yyy-2\OO\times\UU \nonumber \\
&&-\cs^2\nab\ln\rho+\frac{1}{\rho}\left[\JJ\times\BB+\nab\cdot(2\rho\nu\SSSS)\right],
\label{dUdt}
\end{eqnarray}
where $\DDD/\DDD t=\partial/\partial t+\VV\cdot\nab$ is the advective
derivative with respect to the shear flow, $\rho$ is the density,
$\BB$ is the magnetic field, $\JJ=\nab\times\BB/\mu_0$ is the current density,
$\mu_0$ is the vacuum permeability, $\nu$ is the viscosity,
and $\SSSS$ is the traceless rate-of-strain tensor with the components
${\sf S}_{ij}=(\partial_i U_j+\partial_j U_i)/2-\delta_{ij}\nab\cdot\UU/3$.
The tensor $\SSSS$ is not to be confused with the constant scalar $S$,
which denotes the uniform background shear when $q\neq0$.
The logarithmic density obeys the continuity equation in the form
\begin{equation}
\frac{\DDD\ln\rho}{\DDD t}=-\UU\cdot\nab\ln\rho-\nab\cdot\UU.
\end{equation}
The magnetic field $\BB=\nab\times\AAA$ is solved in terms of the
magnetic vector potential $\AAA$,
\begin{equation}
\frac{\DDD\AAA}{\DDD t}=-SA_y\xxx-\EE-\nab\Phi,
\label{dAdt}
\end{equation}
where $\EE=\eta\mu_0\JJ-\UU\times\BB$ is the electric field, with
$SA_y\xxx\equiv\EE_\mathrm{S}$ being the contribution from the shear,
$\eta$ is the magnetic diffusivity, and $\Phi$ is the electrostatic potential.
In \Eq{dAdt}, we have adopted the advecto-resistive gauge, in
which $\Phi=-V_y A_y-\eta\nab\cdot\AAA$ \citep{Can+11}.
As shown in \cite{BNST95}, the inclusion of the advective term $V_y
A_y$ is necessary for being able to adopt shearing--periodic boundary
conditions.
This means that the magnetic diffusion operator reduces to
$\eta\nabla^2\AAA$; see \cite{Can+11} for further details.
In some cases, we also compute the vector potential in the
Coulomb gauge, $\AAA^\mathrm{Cou}=\AAA-\nab\Lambda$, by solving
$\nabla^2\Lambda=\nab\cdot\AAA$ with appropriate boundary conditions.

The nonhelical forcing function $f(\xx,t)$ is given by \citep{HBD04}
\begin{equation}
f(\xx,t)=f_0 \cs \, (\cs k/\delta t)^{1/2}
\frac{\ee\times\kk}{|\ee\times\kk|}
\,e^{\ii\kk(t)\cdot\xx+\ii\varphi(t)},
\end{equation}
where $\ee$ is a random vector that is not aligned with $\kk$,
$\varphi(t)$ is a random phase ($|\varphi|\leq\pi$), and $f_0$ is the
amplitude.
At each time step, a new forcing vector $\kk(t)$ is chosen randomly from
a set of wavevectors $\kk$ whose lengths $|\kk|$ lie in a narrow band
$\kf-\delta k/2\leq k< \kf+\delta k/2$, where $\kf=8$ and $\delta k=1$
is used for all the runs discussed in this paper.

To analyze the possibility of large-scale dynamo action, it is useful
to compute planar averages.
Owing to the inhomogeneity in the $z$-direction, we adopt
$xy$ averages, which are denoted by an overbar, e.g.,
$\meanUU(z,t)=\int\uu(\xx,t)\,\dd x\,\dd y/L^2$.
Fluctuations about the average are then denoted by lowercase symbols,
e.g., $\uu=\UU-\meanUU$, $\bb=\BB-\meanBB$, and $\jj=\JJ-\meanJJ$.

\subsection{Control parameters and initial conditions}
\label{sec:parameters}

The value of the overall root-mean square (rms) velocity $\urms$ is
characterized by the Mach number, $\Ma=\urms/\cs$.
When there is shear, the value of $\urms$ does not include this shear flow.
Since we are here not interested in studying compressibility effects,
we adopt subsonic Mach numbers and take $\Ma\la0.1$ for all runs.
The values of $\nu$ and $\eta$ are characterized by the fluid and
magnetic Reynolds numbers,
\begin{equation}
\Rey=\urms/\nu\kf\quad\mbox{and}\quad
\Rm=\urms/\eta\kf,
\end{equation}
respectively.
The ratio $\Pm=\nu/\eta$ is the magnetic Prandtl number.
In the following, we vary $\Pm$ by keeping the value of $\Rey$ fixed.
Another control parameter is the relative forcing wavenumber, $\kf/k_1$.
The amount of rotation and shear are quantified by the Coriolis and
shear numbers,
\begin{equation}
\Co=2\Omega/\urms\kf\quad\mbox{and}\quad
\Shear=S/\urms\kf,
\end{equation}
respectively.
As initial condition, we use $\UU=\ln\rho/\rho_0=0$, so the initial density
is equal to some reference density $\rho_0$.

The initial magnetic vector potential is calculated from a weak
Gaussian-distributed field with an rms value $\Brms$ such that the
rms Alfv\'en speed $\vA=\Brms/\sqrt{\mu_0\rho_0}$ is a small fraction
of $\cs$.
When $\Rm$ exceeds a certain critical value, there is dynamo action,
i.e., $\vA/\cs$ grows exponentially and saturates eventually at a value
around 0.1.
Instead of quantifying $\vA/\cs$, it is useful to quantify the ratio
$\vA/\urms$, or, equivalently, the value of $\Brms$ in units of the
equipartition field strength, $\Beq=\sqrt{\mu_0\rho_0}\,\urms$.
The rms value of the large-scale field is denoted by $\meanBrms$.

Owing to the presence of rotation and stratification of the
turbulent intensity, we expect the generation of kinetic helicity,
$\overline{\oo\cdot\uu}$, where $\oo=\nab\times\uu$ is the vorticity of
the fluctuating velocity.
Following \cite{Jab+14}, we determine the resulting kinetic helicity in
terms of the nondimensional ratio
\begin{equation}
\epsf(z)=\overline{\oo\cdot\uu}/\urms^2\kf,
\end{equation}
which is characterized primarily by the amplitude of its variation,
defined here as
\begin{equation}
\epsfz=2\bbra{\epsf(z)\sin k_1 z}.
\end{equation}
As in \cite{Jab+14}, we expect $\epsfz$ to increase linearly with
increasing rotation rate and with increasing stratification of turbulent
intensity, provided these values are not too large.

The simulations are performed with the \textsc{Pencil Code} \citep{PC}.
Numerical results are usually presented as averages over a statistically
steady stretch in time.
Error margins are estimated as the largest departure over any one third
of the full time series.

\subsection{Quasi-kinematic test-field method} \label{sec:TFM}

To characterize the nature of large-scale dynamo action, we need to obtain
the mean-field dynamo coefficients that characterize the dependence of
the mean electromotive force $\meanEMF$ on $\meanBB$ and $\meanJJ$.
The most accurate choice is the test-field method
\citep{Sch05,Sch07,Bra05QPO,BRS08}.

In the test-field method, we solve the equations for the fluctuations
$\bbT$ that result from a certain test field $\BBT$.
We represent it by $\bbT=\nab\times\aaT$ and solve for the vector
potential $\aaT$, which, in the Weyl (or temporal) gauge with zero
electrostatic potential, obeys
\begin{equation}
\frac{\partial\aaT}{\partial t}=\uu\times\BBT+\meanUU\times\bbT
+\uu\times\bbT-\overline{\uu\times\bbT}+\eta\nabla^2\aaT.
\label{TFeqn}
\end{equation}
This allows us to compute $\meanEMF^\mathrm{T}=\overline{\uu\times\bbT}$.
We adopt the parameterization $\overline{\cal E}_i^\mathrm{T}=\alpha_{ij}
\overline{B}_j^\mathrm{T}-\eta_{ij}\mu_0\overline{J}_j^\mathrm{T}$.
Since only the $x$- and $y$-components are significant, we have 8 unknowns
for the 4 components of $\alpha_{ij}$ and the 4 components of $\eta_{ij}$.
To obtain all unknowns, we use the 4 vectorial test fields $\BBT=(c,0,0)$,
$(s,0,0)$, $(0,c,0)$, and $(0,s,0)$, where $c=\cos\kT z$ and $s=\sin\kT z$.
Since only the $x$- and $y$-components are significant, we have exactly
8 independent equations for the 8 unknowns.
We choose $\kT=k_1$ and refer to \cite{BRS08} regarding the
significance of also studying $\kT>k_1$ to obtain full integral kernels in
a parameterization involving integral kernels.

When $\uu$ in \Eq{TFeqn} is a solution of the nonlinear \Eq{dUdt}
with the Lorentz force included, we talk about the quasi-kinematic
test-field method.
This method is nonlinear in the sense that it describes correctly the
modifications of the velocity field in response to the actual magnetic
field in the simulations \citep{BRRS08,Karak+14}.
However, it is not fully nonlinear in the sense that it does not
include the fluctuating magnetic field from a small-scale dynamo
\citep{RB10,Kapy+22}.
On the other hand, there are so far no clear cases of practical interest
where the quasi-kinematic method is known to fail; see \cite{Bran18}
for a review.
The only exception is the case where magnetic fluctuations are produced
by applying externally maintained currents to drive the system.
Those cases are mainly of academic interest and not relevant to our
problem at hand.
The success of the quasi-kinematic method lies probably in the fact that
the small-scale dynamo-generated magnetic fields are not well correlated
with the large-scale field.

In the present case, the time-averaged turbulent transport coefficients
depend on $z$.
In addition to plotting the individual components of $\alpha_{ij}$
and $\eta_{ij}$, we also compute the traces
$\alpha\equiv(\alpha_{xx}+\alpha_{yy})/2$ and
$\etat\equiv(\eta_{xx}+\eta_{yy})/2$, as well as the antisymmetric parts,
$\gamma\equiv(\alpha_{yx}-\alpha_{xy})/2$ and
$\delta\equiv(\eta_{xy}-\eta_{yx})/2$.
In the following, we fit $\etat$ to Legendre polynomials of $\cos k_1 z$.
Since these are orthogonal polynomials, a decline of the coefficients with
increasing order can be interpreted as convergence.
Another quantity of interest is the $z$-dependent dynamo number,
$C_\alpha=\alpha/\etat k_1$.
In the present case, it turns out that, to a good approximation, it has
a linear profile.
The values and slopes are given in tabular form below.

\subsection{Mean-field evolution} \label{sec:MFM}

To assess the importance of the aforementioned turbulent transport
coefficients $\alpha$, $\etat$, $\gamma$, and $\delta$, we consider
numerical mean-field models, where we can rescale the coefficients to
learn about their relative importance.
In that case, we solve the one-dimensional mean-field equation, again in the
Weyl gauge,
\begin{equation}
\frac{\partial\meanAA}{\partial t}=\alpha\meanBB
-\etaT\mu_0\meanJJ
+\ggamma\times\meanBB
+\ddelta\times\mu_0\meanJJ
-S\meanA_y\xxx,
\label{MFE}
\end{equation}
where $\ggamma=(0,0,\gamma)$ and $\ddelta=(0,0,\delta)$ are vectors that
only have a $z$-component and $\etaT=\etat+\eta$ is the total magnetic
diffusivity.
It should be remembered, however, that the values and profiles of the
turbulent transport coefficients have been computed under the assumption
of steady mean fields.
This is obviously not the case; see \cite{HB09} for the treatment
of time-dependent mean fields.

\subsection{Magnetic helicity fluxes} \label{sec:helfluxes}

The saturation level of the resulting mean magnetic field is known to
be severely limited by the ability to shed magnetic helicity from
the dynamo through magnetic helicity fluxes; see \cite{Zhou+Blackman24}
for a recent assessment.
It is therefore of interest to consider the evolution equation
for the magnetic helicity balance separately for the large-scale
and small-scale contributions by splitting the total mean
magnetic density, $\overline{\AAA\cdot\BB}$, into two parts:
$\meanAA\cdot\meanBB$ and $\overline{\aaaa\cdot\bb}$.
The evolution equation for $\overline{\AAA\cdot\BB}$ is obtained by
dotting \Eq{dAdt} with $\BB$ and adding its curl dotted with $\AAA$, 
which yields
\begin{equation}
\frac{\partial}{\partial t}\overline{\AAA\cdot\BB}=
-2\overline{\EE\cdot\BB}-\nab\cdot\meanFFFF,
\end{equation}
where $\meanFFFF=\overline{\EE\times\AAA}+\overline{\Phi\BB}$ is the
total magnetic helicity flux.
Note also that $\overline{\EE\cdot\BB}=\eta\nu_0\overline{\JJ\cdot\BB}$,
i.e., the induction term does not contribute.
The evolution equation for $\meanAA\cdot\meanBB$ is obtained from
the evolution equation for the mean field $\partial\meanAA/\partial t
=-\meanEE-\nab\meanPhi$, where
\begin{equation}
\meanEE=\eta\mu_0\meanJJ-\meanUU\times\meanBB-\meanEMF
\label{EmeanDef}
\end{equation}
is the averaged electric field, and $\meanEMF=\overline{\uu\times\bb}$
is the mean electromotive force from the fluctuating fields.
This yields
\begin{equation}
\frac{\partial}{\partial t}\meanAA\cdot\meanBB=
2\meanEMF\cdot\meanBB-2\eta\mu_0\meanJJ\cdot\meanBB
-\nab\cdot\meanFFFF_\mathrm{m},
\label{FluxLS}
\end{equation}
where $\meanFFFF_\mathrm{m}=\meanEE\times\meanAA+\meanPhi\,\meanBB$
is the magnetic helicity flux from the mean field.
Note, however, that in our cases $\meanB_z=0$ at all times owing to the
fact that $\nab\cdot\meanBB=0$, the use of planar averages, and the fact
that $\meanB_z=0$ initially.
Therefore, $\meanPhi\,\meanBB=0$.
Furthermore, the mean-field shear contribution, $\meanEE_\mathrm{S}$,
to the mean electric field only leads to a lateral magnetic helicity
flux and is therefore irrelevant.

Finally, the evolution equation for $\overline{\aaaa\cdot\bb}$ is obtained
from the difference $\overline{\AAA\cdot\BB}-\meanAA\cdot\meanBB$,
which yields
\begin{equation}
\frac{\partial}{\partial t}\overline{\aaaa\cdot\bb}
=-2\meanEMF\cdot\meanBB-2\eta\mu_0\overline{\jj\cdot\bb}
-\nab\cdot\meanFFFF_\mathrm{f},
\label{FluxSS}
\end{equation}
where $\meanFFFF_\mathrm{f}=\overline{\ee\times\aaaa}+\overline{\phi\bb}$
is the magnetic helicity flux of the fluctuating field, $\ee=\EE-\meanEE$
is the fluctuating electric field, and $\phi=V_y a_y-\eta\nab\cdot\aaaa$.
Contrary to \cite{Bra18}, we use here the more natural and more compact
notation where $\meanEMF$ is included in the definition of $\EE$;
see \Eq{EmeanDef}.

\begin{table*}[t!]\caption{
Summary of the results for our test-field runs.
Run~E has uniform shear with $q=0.5$.
}\vspace{12pt}\center{\begin{tabular}{ccccccccccccccccc}
    &          &        &       &       &                & \multicolumn{4}{c}{------ $\quad\etatz^{-1}\times\quad$ ------} & & & & & & \multicolumn{2}{c}{--- $\Beq^{-1}\times$ ---} \\
Run & $\Shear$ & $\Rey$ & $\Rm$ & $\Pm$ & $\eta k_1/\cs$ & $\etat^{(0)}$ & $\etat^{(1)}$ & $\etat^{(2)}$ & $\etat^{(3)}$ & $C_\alpha(z)$ & $-\epsfz$ & $\epsfzM$ & $k_\mathrm{eff}$
& $\Co$ & $\Brms$ & $\meanBrms$ \\
\hline
A &  0   & 17.5 &  3.5 & 0.2 & $2.5\times10^{-3}$ & 0.80 & 0.89 & 0.17 & 0.01 & $2.81\,z$ & 0.48 &$\!-0.06$& 8.8 & 1.78 & 0.52 & 0.43 \\
B &  0   & 16.9 & 16.9 &  1  &   $5\times10^{-4}$ & 0.71 & 0.87 & 0.24 & 0.06 & $2.52\,z$ & 0.47 &$\!-0.12$& 11  & 1.84 & 0.57 & 0.34 \\
C &  0   & 16.4 & 82   &  5  &          $10^{-4}$ & 0.71 & 0.87 & 0.24 & 0.06 & $2.30\,z$ & 0.48 &$\!-0.14$& 14  & 1.91 & 0.58 & 0.22 \\
D &  0   & 15.8 &158   & 10  &   $5\times10^{-5}$ & 0.64 & 0.76 & 0.18 & 0.04 & $2.22\,z$ & 0.47 &$\!-0.14$& 17  & 1.98 & 0.59 & 0.19 \\
\hline
E & 0.63 & 12.3 &123   & 10  &   $5\times10^{-5}$ & 0.49 & 0.45 & 0.01 & 0.04 & $1.37\,z$ & 0.22 &    0.007& 3.0 & 2.50 & 1.39 & 0.34 \\
\label{Tsummary}\end{tabular}}\end{table*}

In the statistically steady state, we can drop the time derivative.
Instead of considering volume-integrated quantities separately for the
northern and southern hemispheres, it is convenient to plot them as
fluxes in terms of undetermined integrals,
\begin{equation}
\int_{z_-}^z 2\eta\mu_0\meanJJ\cdot\meanBB\,\dd z=+
\int_{z_-}^z 2\meanEMF\cdot\meanBB\,\dd z-\meanFFFmz,
\label{LSflux}
\end{equation}
\begin{equation}
\int_{z_-}^z 2\eta\mu_0\overline{\jj\cdot\bb}\,\dd z=-
\int_{z_-}^z 2\meanEMF\cdot\meanBB\,\dd z-\meanFFFfz,
\label{SSflux}
\end{equation}
in the range $z_-\leq z\leq z_+$, where $z_\pm=\pm\pi/k_1$ are the upper
and lower boundaries of the cube.
In the following, we refer to these as ``integrated terms.''

Note that in both \Eqs{LSflux}{SSflux} there are three terms of which
two are manifestly gauge invariant.
Therefore, the third term also, $\meanFFFmz$ and $\meanFFFmz$ in each
equation, respectively, must be gauge invariant.
This argument was already applied by \cite{HB10} in their work on magnetic
helicity fluxes from a dynamo embedded in a conducting halo.

It is convenient to present the magnetic helicity fluxes in normalized form.
For the following, we define the reference flux as
\begin{equation}
F_\mathrm{m0}=\etatz k_1\int_{z_-}^{z_+}\meanBB^2\,\dd z.
\label{RefFlux}
\end{equation}
This is analogous to the work of \cite{Bra18}, except that there
$k_1^2$ instead of $k_1$ was written by mistake.

To compare the current helicity with the kinetic helicity, we define
$\epsf^\mathrm{M}(z)=\overline{\jj\cdot\bb}/\urms^2\kf$ and its amplitude
as $\epsfz^\mathrm{M}=2\bbra{\epsf^\mathrm{M}(z)\sin k_1z}$.
Finally, the ratio between the small-scale current and
magnetic helicity densities is characterized by the ratio
$k_\mathrm{eff}^2=\overline{\jj\cdot\bb}/\overline{\aaaa\cdot\bb}$,
where the two terms have been computed from a sinusoidal fit, analogously
to $\epsfz^\mathrm{M}$.
These quantities are discussed in \Sec{sec:Results}.
It should be noted, however, that the departure of $k_\mathrm{eff}$ from
the value of $\kf$ is mainly a measure of the departure of the magnetic
vector potential from the Coulomb gauge, because the term $\nab\cdot\AAA$
in the expression for $\JJ=-\nabla^2\AAA+\nab\nab\cdot\AAA$ can be
important.
By contrast, the ratio $\overline{(-\nabla^2\aaaa)\cdot\bb}/
\overline{\aaaa\cdot\bb}$ is typically found to be close to the actual
value of $\kf=8\,k_1$, even when the Coulomb gauge is not used.

\begin{figure}[t!]
\plotone{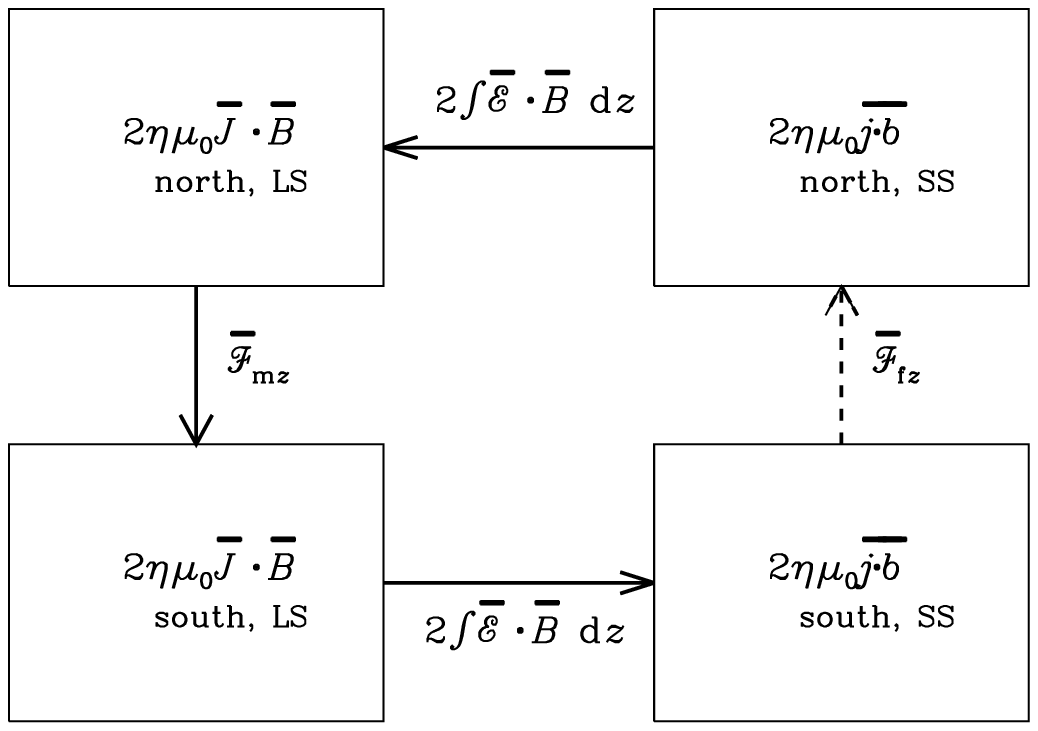}
\caption{
Sketch of the magnetic helicity fluxes between north and south (upper
and lower boxes), and between large scales (LS, left) and small scales
(SS, right).
In the steady state, the four magnetic helicity reservoirs can still
have sinks or sources because of the microphysical resistivity.
This can still be important, especially at small scales, and therefore
the small-scale magnetic helicity fluxes, $\meanFFFF_{\mathrm{f}z}$,
may not balance the $2\meanEMF\cdot\meanBB$ term, which is therefore
indicated by the dashed arrow.
\label{fig:psketch}}
\end{figure}

\subsection{Magnetic helicity cycle} \label{sec:helfluxes2}

In this section, we explain that there is a continuous flux of magnetic
helicity both between hemispheres and between scales.
This is illustrated in \Fig{fig:psketch}.

Owing to the presence of rotation and a finite gradient in the turbulent
intensity, an $\alpha$ effect is expected based on the formula by
\cite{SKR66},
\begin{equation}
\alpha\approx-\ell^2\OO\cdot\nab\ln(\rho\urms),
\label{alphaFormula}
\end{equation}
where $\ell$ is a suitable length scale.
We refer here to \cite{RK93} for analytical calculations based on the
consideration of homogeneous background turbulence that is being affected
by stratification and rotation, and \cite{Bran+13} for simulation results
over a broad range of astrophysical settings.
\Eq{alphaFormula} predicts a positive (negative) value of $\alpha$
in the upper (lower) disk plane.
This, in turn, implies negative (positive) kinetic helicities of the
small-scale velocity and magnetic fields in the upper (lower) disk plane.
It is known that, at least in the absence of shear, the magnetic helicity
of the small-scale field is then also negative (positive) in the upper
(lower) disk plane \citep{DSGB13, Rin21}.
Assuming that small-scale magnetic helicity is transported down the
gradient of the magnetic helicity density and/or magnetic energy density
\citep{KR22}, we expect a small-scale magnetic helicity flux from south
to north; see the dashed line in \Fig{fig:psketch}.
It is shown here as a dashed line, because in our simulations without
shear this flux appears to be too weak, while in our runs with shear
it appears to be too large to balance the corresponding fluxes at
large scales.

There is also a magnetic helicity flux from small to large scales,
which is given by the integral of $2\meanEMF\cdot\meanBB$.
This term has a contribution $\alpha\meanBB^2$, which is positive in
the north, but since it enters with a minus sign, the associated flux
points from small to large scales, and in the opposite direction in the
south where $\alpha$ is negative; see \Fig{fig:psketch}.

\begin{figure*}[t!]
\plotone{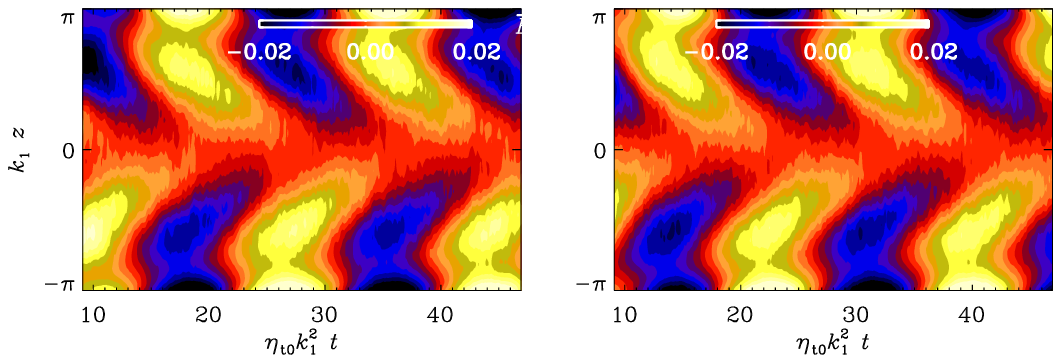}
\caption{
Butterfly diagrams for $\meanB_x$ (left) and $\meanB_y$ (right)
for Run~D with $\Pm=10$, $\eta=5\times10^{-5}$, $\nu=5\times10^{-4}$.
\label{fig:ppbutter_E256a_5em5b}}
\end{figure*}

\begin{figure*}[t!]
\plotone{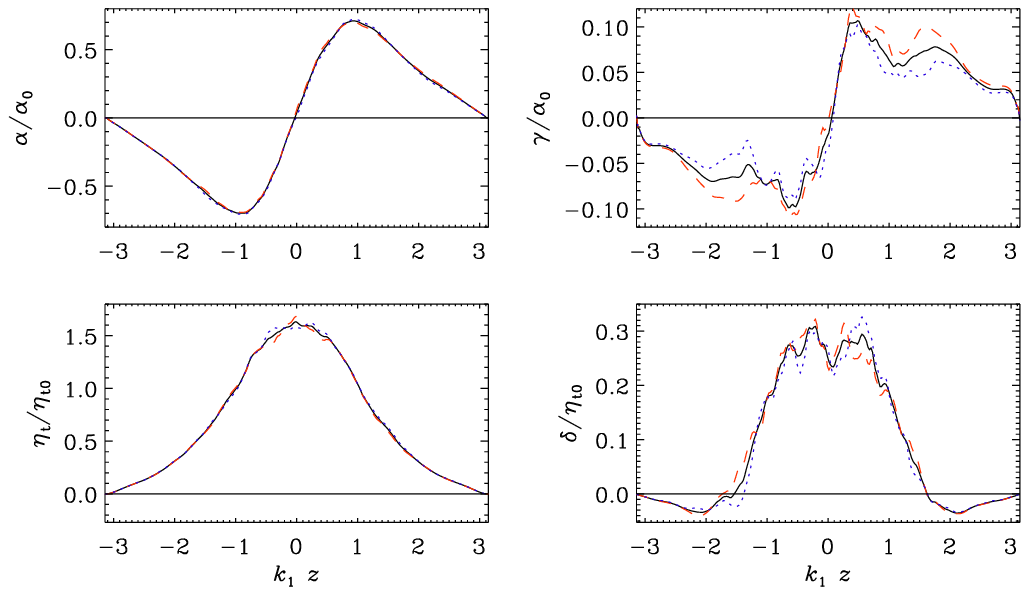}
\caption{
The black lines denote the time-averaged normalized profiles of
$\alpha$, $\gamma$, $\etat$, and $\delta$ for Run~D with $\Pm=10$,
$\eta=5\times10^{-5}$, $\nu=5\times10^{-4}$.
In the four panels, the red (blue) lines denote $\alpha_{xx}$
($\alpha_{yy}$), $\alpha_{yx}$ ($-\alpha_{xy}$), $\eta_{xx}$
($\eta_{yy}$), and $\eta_{xy}$ ($-\eta_{yx}$).
The ratio, $\alpha/\etat k_1$, shows local extrema at $k_1 z=\pm2$
of about 5, but has here a nearly linear profile as a function of $z$.
\label{fig:ppalp_TE256e}}
\end{figure*}

\section{Results}
\label{sec:Results}

We have performed a series of runs with different parameters; see
\Tab{Tsummary} for a summary.
In all cases without shear, we use $256^3$ mesh points and keep the
viscosity fixed ($\nu k_1/\cs=5\times10^{-4}$), so the level of turbulence
stays unchanged but $\Rm$ is increased by decreasing the value of $\eta$.
This makes the dynamo stronger, so $\Brms/\Beq$ increases.
This quenches the velocity field, and therefore the values of $\Rey$
decrease with increasing values of $\Pm$.
We notice, however, that as $\Pm$ and $\Brms/\Beq$ increase, the
normalized rms value of the mean field, $\meanBrms/\Beq$, decreases.
As already emphasized above, this is at the core of the problem of
catastrophic quenching.

For Runs~A--E, we also have obtained test-field results,
while for Runs~F--H, we have focussed on the analysis of
the contributions to the magnetic helicity balance.
In some of those runs, we also increased the resolution.
Those runs have uniform shear with $q=0.5$.
This choice is motivated by demanding that $|\VV|/\cs$ does not exceed unity.
Given that the value $q=3/2$ is of particular interest for accretion
disks, we also consider one such case using then, however, a correspondingly
larger sound speed of $\cs=3$ so that $|\VV|/\cs$ is unchanged.
The values of $\Rey$ and $\Rm$ are given for the saturated state.

\subsection{Dynamos without shear} \label{sec:WithoutShear}

For all the cases in \Tab{Tsummary}, there is dynamo action.
The resulting magnetic field has spatiotemporal coherence with
a systematic large-scale magnetic field oscillation on a time
scale long compared with the turnover time of the turbulence.
As noted above, however, although $\Brms$ is seen to increase with
increasing magnetic Reynolds number, $\Rm=\Pm\Rey$, the rms magnetic
field contained in the mean field, $\meanBrms$, is seen to decrease.

In \Fig{fig:ppbutter_E256a_5em5b}, we show $zt$ diagrams, also known
as butterfly diagrams, for the two relevant components of $\meanBB$
for Run~D with $\Pm=10$.
We see migration of the magnetic field away from the boundaries at
$k_1 z_\pm=\pm\pi$.
This is expected for $\alpha^2$ dynamos with a nonuniform distribution
of $\alpha$ \citep{Stefani+Gerbeth03, Cole+16}, or even for a uniform
$\alpha$, but with different boundary conditions on two opposite ends;
see \cite{Bra17}.

\subsection{Turbulent transport coefficients} \label{sec:TransportCoefficients}

To identify the nature of the large-scale dynamo seen above, it is
useful to compute the turbulent transport coefficients for the horizontal
averages applied in this study.
All our test-field runs have $\nu=5\times10^{-4}\cs/k_1$.
We adopt the expansion
$\etat=\sum_{\ell=0}^{3}\etat^{(\ell)} P_\ell(\cos k_1z)$.
In the range $k_1 |z|\leq 2.5$, we determine a linear fit to
$C_\alpha(z)\approx\alpha/\etat k_1$.
For larger values of $|z|$, $C_\alpha$ varies no longer linearly, so
this part is ignored in the fit.

In \Fig{fig:ppalp_TE256e}, we show time-averaged profiles of $\alpha(z)$,
$\gamma(z)$, $\etat(z)$, and $\delta(z)$ for Run~D with $\Pm=10$.
For normalization purposes, we adopt the following estimates
for the $\alpha$ effect and the turbulent magnetic diffusivity:
\begin{equation}
\alpha_0=-\epsfz\urms/3,\quad
\etatz=\urms/3\kf.
\end{equation}
Note that in all our simulations (both with and without shear) $\alpha_0$
is negative.
In \Fig{fig:ppalp_TE256e}, the ratio $\alpha/\etat k_1$ shows local
extrema at $k_1 z=\pm2$ of about $\pm5$, but has here a more linear profile
as a function of $z$ than for Run~A.
The profile $\gamma(z)$ quantifies turbulent pumping.
It is negative in the southern hemisphere and positive in the northern
hemisphere.
Since $\ggamma$ plays the role of an advection vector (albeit without
any material motion), this corresponds to magnetic field pumping away
from the midplane; see \cite{SS22} for a review.
The profile $\delta(z)$ describes the $\OO\times\meanJJ$ or R\"adler
effect \citep{Radler69}, which is known to contribute to dynamo action,
although this term alone cannot contribute to a change in $\meanBB^2$.
The sign of $\delta$ is here as expected from theory, and it also
agrees with earlier test-field results \citep{BRRK08}.

For the corresponding results for Run~A with $\Pm=0.2$, the ratio
$\alpha/\etat k_1$ again shows local extrema at $k_1 z=\pm2$ of about
$\pm5$.

\subsection{Importance of mean-field effects}

To assess the relative importance of the turbulent transport coefficients,
we have solved the relevant mean-field model with \Eq{MFE} using the
coefficient from Run~D.
We find that the model with all the mean-field transport coefficients
included yields a slow growth with the growth rate $\lambda/\cs k_1=0.0021$.
The fact that this number is different from zero, even though the
original direct numerical simulation has reached a steady state,
remains unexplained.
To reach a marginally excited state, we would need to scale down the
$\alpha$ tensor by a factor of about four to reach a marginally excited
state.
Similar departures from the expected vanishing growth rate have been
seen before; see \cite{Warnecke+21} for simulations in spherical geometry,
where the $\alpha$ tensor needed to be scaled up to reach a marginally
excited state.

\begin{table}[t!]\caption{
Growth rates $\lambda$ for mean-field models for different combinations
of $c_\gamma$ and $c_\delta$ using the test-field results for Run~D,
Run~D with shear, and Run~E with shear.
In the last row, the factor $c_\alpha$ by which $\alpha$ needs to be
scaled to reach a marginally excited state, is given for Run~D.
}\hspace{-10mm}\vspace{12pt}\centerline{\begin{tabular}{cccccc}
           &            &            & \multicolumn{3}{c}{. . . . . . . $\quad\lambda/\cs k_1\quad$ . . . . . . . } \\
$c_\alpha$ & $c_\gamma$ & $c_\delta$ & Run~D & Run~D+Sh & Run~E+Sh \\
\hline
 1   & 1 & 1 & 0.0021 & 0.023 & 0.0051 \\
 1   & 0 & 1 & 0.0017 & 0.021 & 0.0046 \\
 1   & 1 & 0 & 0.0022 & 0.021 & 0.0054 \\
 1   & 0 & 0 & 0.0020 & 0.020 & 0.0050 \\
0.25 & 0 & 0 &   0    &  ...  &  ...   \\
\label{TMFS}\end{tabular}}\end{table}

\begin{figure*}[t!]
\plotone{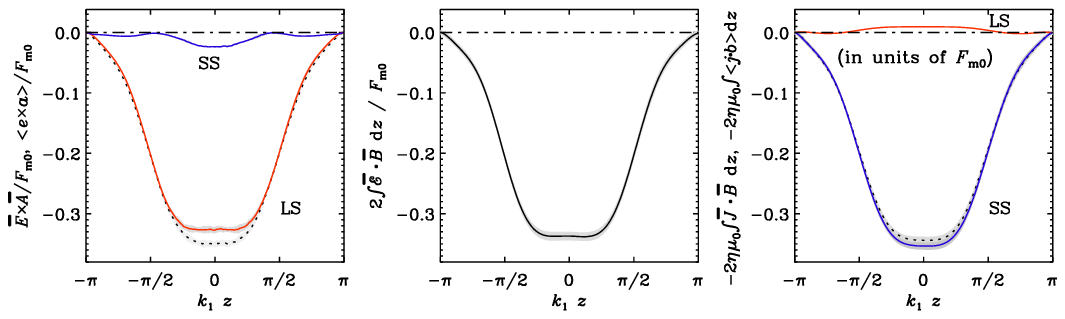}
\caption{
Magnetic helicity fluxes for Run~D with $\Pm=10$,
$\eta=5\times10^{-5}$, and $\nu=5\times10^{-4}$.
The blue (red) lines denote the small-scale (large-scale) contributions,
where applicable, and the black dotted lines denote their sum.
The black dashed--dotted line is the zero line.
Note that the $\overline{\jj\cdot\bb}$ and $\meanJJ\cdot\meanBB$
terms have been plotted with a minus sign, so $2\eta\mu_0 \int
\overline{\jj\cdot\bb}\,\dd z$ is here positive.
\label{fig:pphelflux_E256a_5em5b}}
\end{figure*}

We now study the relative importance of the off-diagonal
terms of both the $\alpha_{ij}$ and $\eta_{ij}$ tensors.
In \Tab{TMFS}, we give the values of $\lambda$ after having
rescaled the off-diagonal components of the two tensors by
scaling coefficients
\begin{equation}
\alpha_{ij}\to c_\gamma \, \alpha_{ij},\quad
\eta_{ij}\to c_\delta \, \eta_{ij},\quad i\neq j.
\end{equation}
Thus, when $c_\gamma=0$ (1), the $\gamma$ effect is ignored (included),
and when $c_\delta=0$ (1), the $\delta$ effect is ignored (included).
For Run~D, we also have shown the factor $c_\alpha=0.25$ by which $\alpha$
needs to be scaled to reach a marginally excited state.
We also studied models in which we included spatiotemporal nonlocality
by solving a differential equation for $\meanEMF$ \citep{RB12, BC18,
Pipin23}, but this did not change the value of $c_\alpha$ significantly.

From the results presented in \Tab{TMFS}, we see that ignoring the
$\gamma$ effect for the profiles from Run~D (\Fig{fig:ppalp_TE256e}),
decreases the growth rate slightly.
Thus, the inclusion of the $\gamma$ effect supports dynamo action
in our case.
On the other hand, ignoring the $\delta$ effect increases the growth
rate slightly.
Therefore, the inclusion of the $\delta$ effect suppresses dynamo action
slightly.
On the other hand, the changes in the growth rate are surprisingly
small, so it is probably reasonable to say that the importance of the
off-diagonal components in the model is small and that the dynamo is
well described by an isotropic $\alpha^2$ dynamo.

Next, we add shear of the same strength as for Run~E (\Tab{TMFS}),
but we continue using the mean-field transport profiles of Run~D
(\Fig{fig:ppalp_TE256e}).
This is obviously inconsistent, but it allows us to isolate the effect of
shear in the mean-field model from that of the profiles for $\alpha_{ij}$
and $\eta_{ij}$.
The overall growth rate is about ten times larger than without shear,
but the differences in the growth rates for different combinations of
$c_\gamma$ and $c_\delta$ are small.
In summary, we find that both the $\gamma$ effect and the $\delta$ effect
contribute slightly to dynamo action, and that excluding them decreases
the growth rate slightly.

On the other hand, when one uses the actual profiles for Run~E together with
shear, we find not only larger differences for different combinations of
$c_\gamma$ and $c_\delta$, but the overall growth rates are
also decreased by a factor of about 4 and are now only about 2.5 times
larger than for the profiles of Run~D and no shear.
In particular, the inclusion of the $\delta$ and shear--current effects
supports dynamo action, while the inclusion of the $\gamma$ effect
diminishes dynamo action.
In units of $\urms\kf$, the value $\lambda/\cs k_1$ corresponds to
$\lambda/\urms\kf=0.01$.

\begin{figure*}[t!]
\plotone{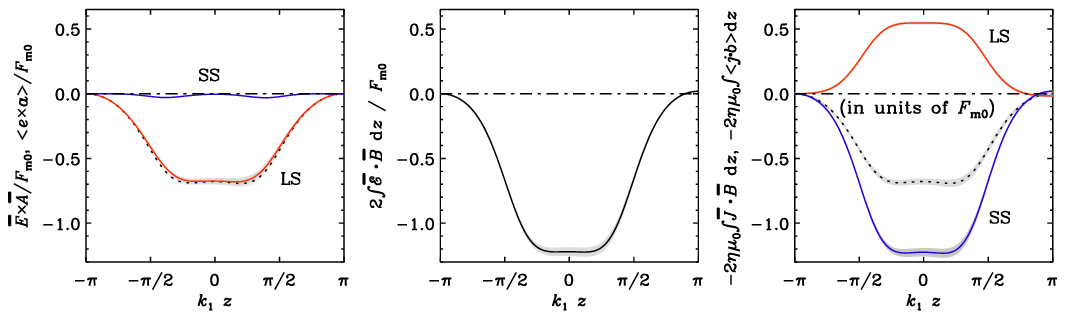}
\caption{
Similar to \Fig{fig:pphelflux_E256a_5em5b}, but for Run~A with $\Pm=0.2$,
$\eta=2.5\times10^{-3}$, and $\nu=5\times10^{-4}$.
Note that the integrated $2\meanEMF\cdot\meanBB$ term is
balanced by the integrated $-2\eta\mu_0\overline{\jj\cdot\bb}$
term, but there is also a significant contribution from the
integrated $2\eta\mu_0\meanJJ\cdot\meanBB$ term, which balances
$\meanEE\times\meanAA$.
\label{fig:pphelflux_TE256d}}
\end{figure*}

\begin{table*}[t!]\caption{
The contributions from $F_\mathrm{m0}^{-1}(\overline{\ee\times\aaaa})_z$,
$2F_\mathrm{m0}^{-1}\int\meanEMF\cdot\meanBB\,\dd z$, and
$-2\eta\mu_0 F_\mathrm{m0}^{-1}\int\overline{\jj\cdot\bb\,\dd z}$
to the magnetic helicity flux balance, along with other properties,
for runs with shear.
}\hspace{-20mm}\vspace{12pt}\centerline{\begin{tabular}{cccccccccccccc}
    & & & &                &  $F_\mathrm{m0}^{-1}\times$ & $-2F_\mathrm{m0}^{-1}\times$       & $\!\!\!-2\eta\mu_0F_\mathrm{m0}^{-1}\times\!\!\!$
    & & & & \multicolumn{2}{c}{--- $\Beq^{-1}\times$ ---} \\
Run & $\Rey$ & $\Rm$ & $\Pm$ & $\eta k_1/\cs$ & $(\overline{\ee\times\aaaa})_z$ & $\int\meanEMF\cdot\meanBB\,\dd z$
    & $\int\overline{\jj\cdot\bb}\,\dd z$ & $-\epsfz$ & $\epsfzM$ & $k_\mathrm{eff}$ & $\Brms$ & $\meanBrms$ & $N^3$ \\
\hline
E & 12.3 & 123 & 10  &   $5\times10^{-5}$ & 0.035 & 0.035 &       0.005  & 0.22 &          0.007& 3.0   & 1.39 & 0.34 &  $256^3$ \\
F & 12.9 & 258 & 20  & $2.5\times10^{-5}$ & 0.045 & 0.015 &       0.030  & 0.22 &          0.05 & 15    & 1.58 & 0.31 &  $512^3$ \\
G & 13.2 & 660 & 50  &          $10^{-5}$ & 0.025 & 0.005 &       0.020  & 0.15 &          0.10 & 24    & 2.62 & 0.37 & $1024^3$ \\
\hline
H & 160  & 160 &  1  &   $5\times10^{-5}$ & 0.075 & 0.160&$\!\!\!-0.085~$& 0.33 &$\!\!\!\!-0.02$&$14\ii$& 0.94 & 0.22 &  $512^3$ \\
I & 340  & 340 &  1  &   $2\times10^{-5}$ & 0.060 & 0.040 &       0.020  & 0.26 &          0.04 & 16~   & 1.10 & 0.28 &  $512^3$ \\
J & 540  & 540 &  1  &          $10^{-5}$ & 0.075 & 0.020 &       0.055  & 0.16 &          0.10 & 16~   & 1.24 & 0.24 &  $512^3$ \\
K & 850  & 850 &  1  &   $5\times10^{-6}$ & 0.060 & 0.015 &       0.045  & 0.08 &          0.07 & 17~   & 1.24 & 0.21 &  $512^3$ \\
\hline
L & 410  & 164 & 0.4 &   $5\times10^{-5}$ & 0.050 & 0.130&$\!\!\!-0.080~$& 0.33 &$\!\!\!\!-0.03$&$~9\ii$& 0.94 & 0.22 &  $512^3$ \\
M & 830  & 166 & 0.2 &   $5\times10^{-5}$ & 0.075 & 0.075&$\!\!\!-0.000~$& 0.33 &$\!\!\!\!-0.00$&$~3~  $& 0.91 & 0.20 &  $512^3$ \\
N &1650  & 165 & 0.1 &   $5\times10^{-5}$ & 0.060 & 0.140&$\!\!\!-0.080~$& 0.33 &$\!\!\!\!-0.03$&$13\ii$& 0.92 & 0.25 &  $512^3$ \\
\hline
O & 690  & 345 & 0.5 &   $2\times10^{-5}$ & 0.120 & 0.020&        0.100  & 0.26 &$\!\!\!\!+0.07$&$12~  $& 1.05 & 0.18 &  $512^3$ \\
P &1700  & 340 & 0.2 &   $2\times10^{-5}$ & 0.070 & 0.050&        0.020  & 0.25 &$\!\!\!\!+0.04$&$13~  $& 1.08 & 0.26 &  $512^3$ \\
\hline
Q & 13.3 & 133 & 10  &   $5\times10^{-5}$ & 0.030 & 0.020 &       0.010  & 0.22 &          0.015&$5.6  $& 1.33 & 0.40 &  $256^3$ \\
R & 31.0 & 310 & 10  &   $5\times10^{-5}$ & 0.040 & 0.020 &       0.010  & 0.00 &          0.007& 6.9   & 1.03 & 0.28 &  $256^3$ \\
\label{Tsummary2}\end{tabular}}\end{table*}

\subsection{Interpretation of magnetic helicity fluxes}
\label{sec:HelicityFluxes}

We are interested in the magnetic helicity flux between the northern
and southern hemispheres.
It is convenient to plot the magnetic helicity flux through any $xy$
plane as a function of $z$.
In \Fig{fig:pphelflux_E256a_5em5b}, we see that for Run~D most of the
total magnetic helicity flux is contained in the large-scale contribution,
$\meanEE\times\meanAA$.
The small-scale magnetic helicity flux is nearly negligible.
The large-scale component is nearly balanced by the fluxes
$2\int_{-\pi}^0\meanEMF\cdot\meanBB\,\dd z$ and
$2\int_0^\pi\meanEMF\cdot\meanBB\,\dd z$ between different scales in
the southern and northern hemispheres.
At small scales, however, almost the entire flux is absorbed by the
ohmic diffusion term $2\eta\mu_0\overline{\jj\cdot\bb}$, which was also
found for most of the earlier work \citep{DSGB13, Rin21}, when $\Rm$
was not yet very large.

\begin{figure*}[t!]
\plotone{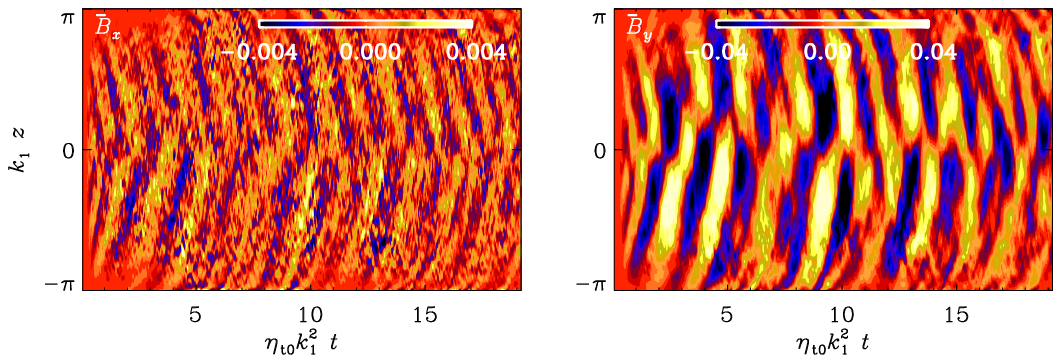}
\caption{
Butterfly diagrams for $\meanB_x$ (left) and $\meanB_y$ (right)
for Run~E with shear and $\Pm=10$.
\label{fig:ppbutter_E256a_5em5_q05b}}
\end{figure*}

For the much more diffusive Run~A with $\Pm=0.2$, the situation
is a bit different; see \Fig{fig:pphelflux_TE256d}.
Here, the integrated $2\meanEMF\cdot\meanBB$ term is still
balanced by the integrated $-2\eta\mu_0\overline{\jj\cdot\bb}$
term, but now there is also a significant contribution from the
integrated $2\eta\mu_0\meanJJ\cdot\meanBB$ term, which balances
$\meanEE\times\meanAA$.
Looking at \Fig{fig:psketch}, this means that the microphysical
magnetic diffusivity is important not only at small scales, when the
integrated $2\meanEMF\cdot\meanBB$ term is entirely supplied by the
integrated $-2\eta\mu_0\overline{\jj\cdot\bb}$ term and not much by the
$\overline{\ee\times\aaaa}$ or integrated $\overline{\phi\bb}$ terms, but
also at large scales, when the integrated $2\meanEMF\cdot\meanBB$ term
is supplied to 50\% by the integrated $\eta\nu_0\overline{\JJ\cdot\BB}$
term and to another 50\% by $\overline{\EE\times\AAA}$.

\subsection{The effect of shear} \label{sec:EffectShear}

We now consider models with finite shear using $q=0.5$ (so $V_y=-q\Omega x$).
This value is less than what is used to model Keplerian shear,
where $q=3/2$.
Nevertheless, one such case will be considered in \Sec{sec:SteeperProfile}.
As emphasized above, this is because we want to avoid supersonic speeds
on the shearing boundaries at $x=\pm\pi$.
In the present case with $\Omega/\cs k_1=0.5$ and $q=0.5$, we have
$V_y(\pm\pi)=\mp0.8\cs$.
Run~E is an example of a model with shear.
The fluxes for this and a few other runs with shear are summarized in
\Tab{Tsummary2}.
The last two cases of Runs~Q and R are designed to explore the
possible sensitivity of the results upon the choice of the shear profile
and its magnitude, and are discussed in \Sec{sec:SteeperProfile}.

\begin{figure*}[t!]
\plotone{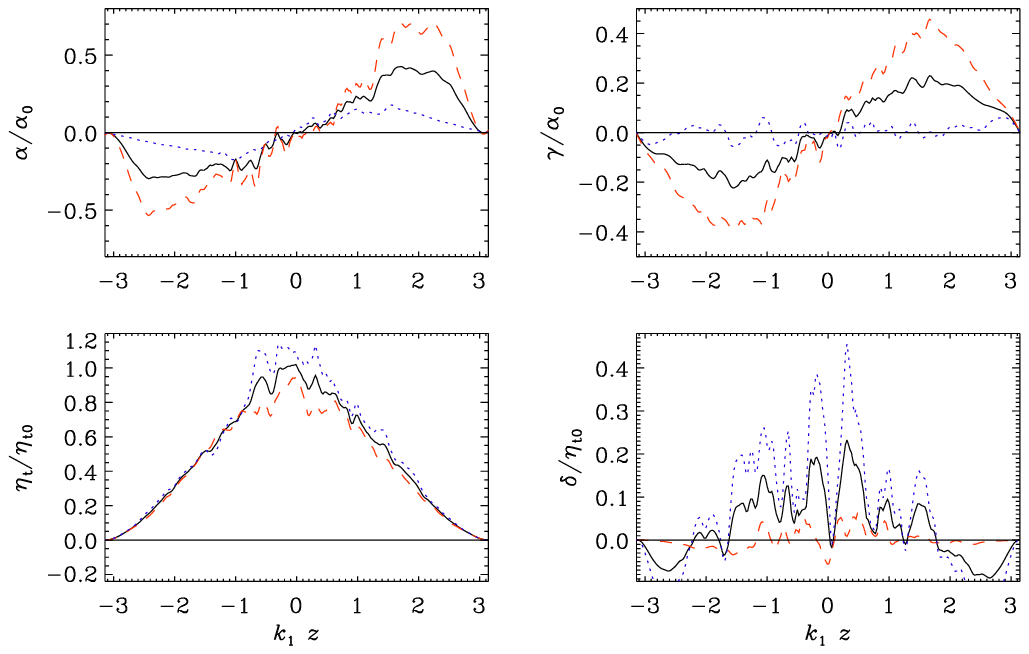}
\caption{
Time-averaged profiles of $\alpha$, $\gamma$, $\etat$, and $\delta$
for Run~E with shear and $\Pm=10$.
The ratio, $\alpha/\etat k_1$, shows local extrema at $k_1 z=\pm2$ of about 5,
but has here a more linear profile as a function of $z$.
The red lines refer to $\alpha_{xx}(z)$, $\alpha_{yx}(z)$,
$\eta_{xx}(z)$, and $\eta_{yx}(z)$, and the blue lines to
$\alpha_{yy}(z)$, $\alpha_{xy}(z)$, $\eta_{yy}(z)$, and $\eta_{xy}(z)$.
Note that, while $\eta_{xx}\approx\eta_{yy}$, we find that $\alpha_{xx}\gg\alpha_{yy}$.
Also, $\alpha_{yx}\gg\alpha_{xy}$, i.e., the pumping of $\meanB_x$ is stronger
than that of $\meanB_y$.
\label{fig:ppalp_TE256a_5em5_q05b}}
\end{figure*}

\begin{figure*}[t!]
\plotone{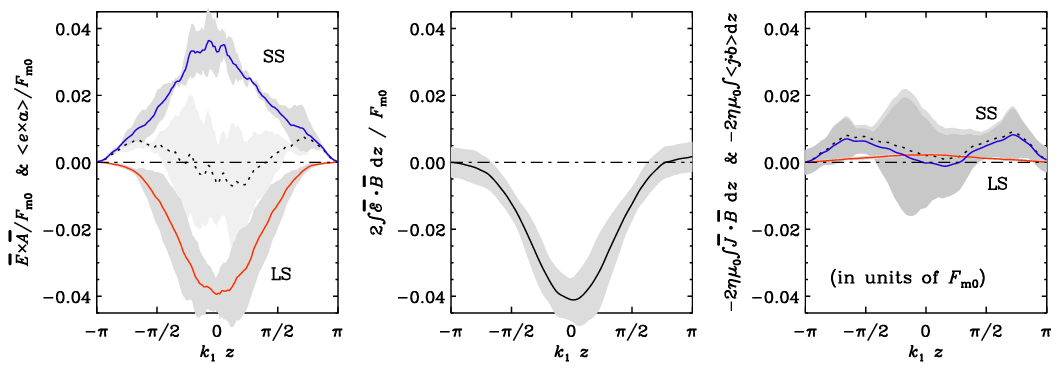}
\caption{
Similar to \Fig{fig:pphelflux_E256a_5em5b}, but for Run~E with shear and $\Pm=10$.
The large-scale and small-scale magnetic helicity fluxes nearly cancel.
In the small-scale magnetic helicity equation, the integrated
$2\meanEMF\cdot\meanBB$ term balances $\overline{\ee\times\aaaa}$,
and the integrated $-2\eta\overline{\jj\cdot\bb}$ term is small.
We recall that the $\overline{\jj\cdot\bb}$ and $\meanJJ\cdot\meanBB$
terms have been plotted with a minus sign, so $2\eta\mu_0 \int
\overline{\jj\cdot\bb}\,\dd z$ is now negative.
\label{fig:pphelflux_E256a_5em5_q05b}}
\end{figure*}

In \Fig{fig:ppbutter_E256a_5em5_q05b}, we show butterfly diagrams
for $\meanB_x$ and $\meanB_y$ for Run~E with shear and $\Pm=10$.
They are consistent with earlier results by \cite{SL90}, where the
field is confined to the disk, which is here accomplished by the use
of perfect conductor boundary conditions; see \cite{BC97} for further
detail and references.
Similarly to the cases without shear, as we increase the value of $\Pm$
further and $\Rm$ increases, the ratio $\Brms/\Beq$ increases, but now the
level of the mean field, $\meanBrms/\Beq$, stays approximately the same.
This might suggest that catastrophic quenching is now alleviated.

\begin{figure*}[t!]
\plotone{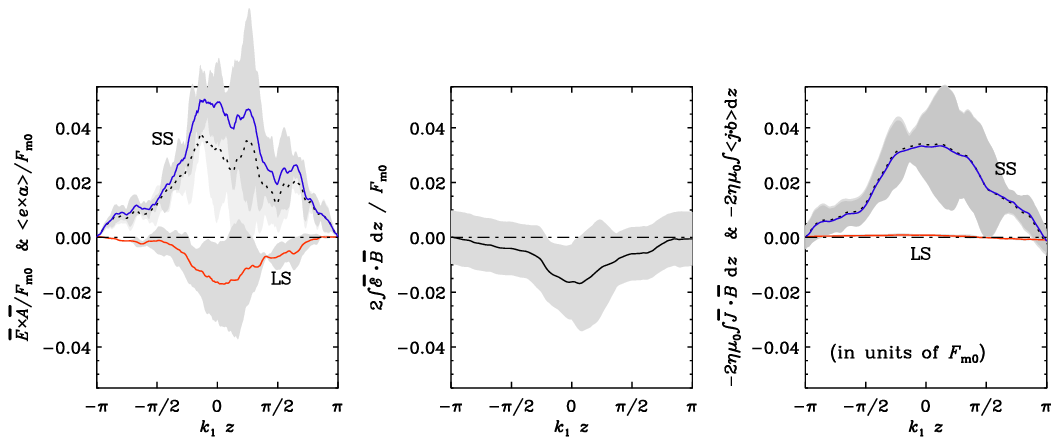}
\caption{
Similar to \Fig{fig:pphelflux_E256a_5em5_q05b}, but for Run~F with $\Pm=20$.
\label{fig:pphelflux_E512a_2p5em5_q05a}}
\end{figure*}

In \Fig{fig:ppalp_TE256a_5em5_q05b}, we show time-averaged profiles of
$\alpha$, $\gamma$, $\etat$, and $\delta$ for Run~E.
We see that, while $\eta_{xx}\approx\eta_{yy}$, we find that
$\alpha_{xx}\gg\alpha_{yy}$.
This agrees with earlier simulations of \cite{Bra05QPO}, but is opposite
to the results of \cite{Gressel+08}.
Furthermore, we find that $\alpha_{yx}\gg\alpha_{xy}$, i.e., the
pumping of $\meanB_x$ is stronger than that of $\meanB_y$; see
\cite{Ossendrijver+02} for earlier work on directionally dependent
pumpings of the magnetic field in a sphere.

The contribution from $\eta_{xy}$ is rather small and, as already
emphasized before, that from $\eta_{yx}$ fluctuates around zero.
It is this component that is relevant for the shear--current effect
\citep{RK03, RK04}.
Its magnetic contribution to the shear--current effect was thought to
be an important driver \citep{Squire+Bhattacharjee15}, but even the
fully nonlinear test-field method did not show such a contribution
\citep{Kapy+20}.

In \Fig{fig:pphelflux_E256a_5em5_q05b}, we show magnetic helicity fluxes
for Run~E.
The large-scale and small-scale magnetic helicity fluxes nearly cancel and
are nearly equally important.
It is striking to note that the small-scale magnetic helicity flux,
arising from the correlated action of the turbulent eddies, roughly
scales as the turbulent energy density and has the same sign throughout
the computational volume.
In the third panel of \Fig{fig:pphelflux_E256a_5em5_q05b}, we see the
integrated small-scale current helicity, which oscillates from zero to
a negative maximum back to zero twice.
This implies that the current helicity, and the associated magnetic
helicity, oscillates through two full cycles within the computational
volume.
The small-scale magnetic helicity flux seen in the first panel shows a
negligible contribution from the turbulent diffusion term of the form
$-\kappat \nabla_z h$, where $\kappat$ is a turbulent diffusivity and
$h=\overline{\aaaa\cdot\bb}$ is the small-scale magnetic helicity density.

\begin{figure*}[t!]
\plotone{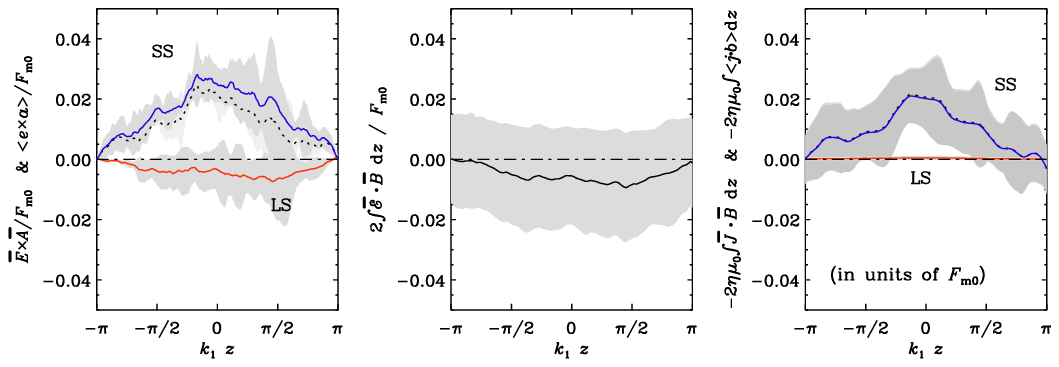}
\caption{
Similar to \Fig{fig:pphelflux_E512a_2p5em5_q05a}, but for Run~G with $\Pm=50$.
\label{fig:pphelflux_E1024a_1em5_q05a}}
\end{figure*}

\begin{figure*}[t!]
\plotone{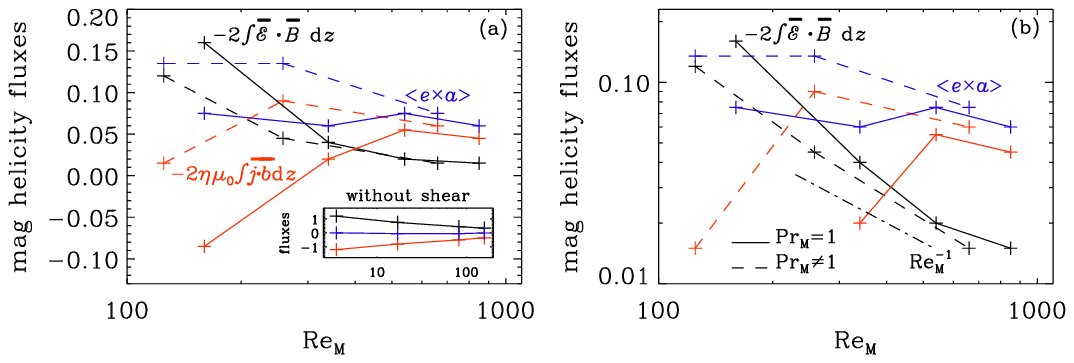}
\caption{
Summary of the small-scale magnetic helicity fluxes (blue line)
and the typical values of $-2\int\meanEMF\cdot\meanBB\,\dd z$ (black
lines) and $-2\eta\mu_0\int\overline{\jj\cdot\bb}\,\dd z$ (red lines)
for the runs with shear in semi-logarithmic (a) and double-logarithmic
(b) representations.
The latter also shows the $\Rm^{-1}$ scaling as the dashed--dotted line.
The dashed lines connecting each three data points are for Runs~E--G
(with $\Pm\neq1$) and the solid lines for Runs~H--K (with $\Pm=1$).
The lines for Runs~E--G have been upscaled by a factor 3 to make them
coincide with those for Runs~H and I.
The inset shows the fluxes for Runs~A--D without shear in the same
color coding.
\label{fig:pfluxes}}
\end{figure*}

In \Eq{FluxLS} for the large-scale helicity equation, the integrated
$2\meanEMF\cdot\meanBB$ term balances $\meanEE\times\meanAA$,
and in \Eq{FluxSS}, also $\overline{\ee\times\aaaa}$ balances
$2\int\meanEMF\cdot\meanBB\,\dd z$, so, contrary to the cases without
shear, the integrated $2\eta\mu_0\overline{\jj\cdot\bb}$ term is small,
and the integrated $2\eta\mu_0\meanJJ\cdot\meanBB$ term is smaller still.
This was not the case in much of the earlier work without shear
\citep{DSGB13, Rin21}.

For Run~F, $\meanEE\times\meanAA$ stays unchanged,
but $2\int\meanEMF\cdot\meanBB\,\dd z$ now decreases and
$2\eta\mu_0\int\overline{\jj\cdot\bb}\,\dd z$ increases and is of opposite
sign compared to before; see \Fig{fig:pphelflux_E512a_2p5em5_q05a}.
This trend persists even for Run~G, although here the statistical
significance is more questionable; see \Fig{fig:pphelflux_E1024a_1em5_q05a}.

The $\Rm$ dependence of the magnetic helicity fluxes in \Tab{Tsummary2}
is unexpected.
This dependence is shown more clearly for Runs~E--G and Runs~H--K in
\Fig{fig:pfluxes}, where $-2\int\meanEMF\cdot\meanBB\,\dd z$ displays
a monotonic decrease proportional to $\Rm^{-1}$.
The small-scale magnetic helicity flux divergence, on the other
hand, is nearly constant in all cases and strongly exceeds
$-2\int\meanEMF\cdot\meanBB\,\dd z$ for large values of $\Rm$.
As a consequence, to obey the steady-state condition of \Eq{FluxSS},
the magnetic helicity dissipation must become important at large $\Rm$.
A similar behavior has not previously been seen in the absence of shear;
see the corresponding plots of \cite{DSGB13} and \cite{Rin21} and the
inset to \Fig{fig:pfluxes}.
Looking again at \Tab{Tsummary2}, we see, however, that the integrated
$\overline{\jj\cdot\bb}$ terms do still obey \Eq{SSflux}, i.e.,
\begin{equation}
\overline{\ee\times\aaaa}=-2\int_{z_-}^z \left(
\meanEMF\cdot\meanBB+\eta\mu_0\overline{\jj\cdot\bb}
\right)\,\dd z.
\label{flux-balance}
\end{equation}

We have seen that for larger values of $\Pm$ and $\Rm$, the small-scale
magnetic helicity flux, $\overline{\ee\times\aaaa}$, stays approximately
unchanged, although the integrated $-2\meanEMF\cdot\meanBB$ term still
declines with larger values of $\Rm$.
Thus, there is now an excess of small-scale magnetic flux from south
to north.
This excess must be dissipated by the integrated ohmic term
$2\eta\mu_0\overline{\jj\cdot\bb}$.
Such a behavior is rather unexpected and it seems to be a general
consequence of dynamos with shear and large values of $\Rm$, and is
not just a specific consequence of large values of $\Pm$.

\begin{figure*}[t!]
\plotone{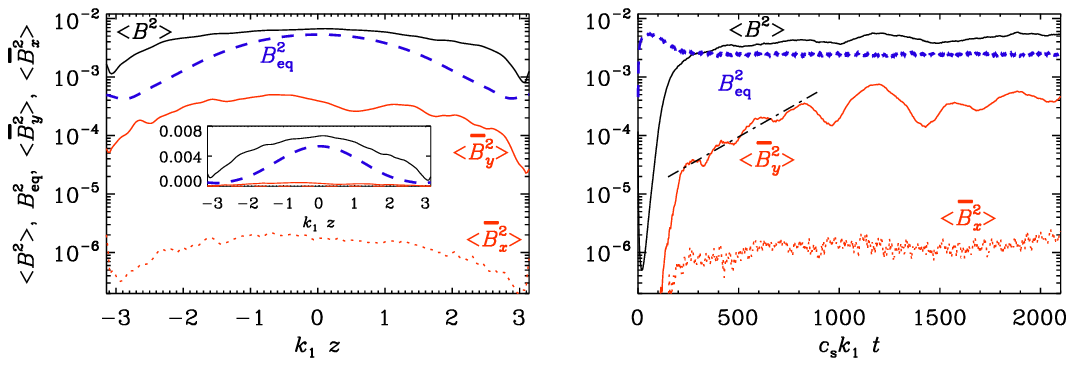}
\caption{
(a) $z$-profiles of $\bra{\BB^2}_{xyt}$ (solid black), $\Beq^2$ (dashed blue),
$\bra{\meanB_y^2}_t$ (solid red), and $\bra{\meanB_x^2}_t$ (dotted red).
(b) $t$-profiles of $\bra{\BB^2}_{xyz}$ (solid black), $\Beq^2$ (dashed blue),
$\bra{\meanB_y^2}_z$ (solid red), and $\bra{\meanB_x^2}_z$ (dotted red)
for Run~E, here plotted in code units, $[B]=\mu_0\rho_0\cs^2$.
The dashed--dotted line indicates a slow but exponential growth with the
growth rate $10^{-2}\,\urms\kf$, for the squared mean-field strength.
\label{fig:pBm_two_E256a_5em5_q05b_rep}}
\end{figure*}

\begin{figure*}[t!]
\plotone{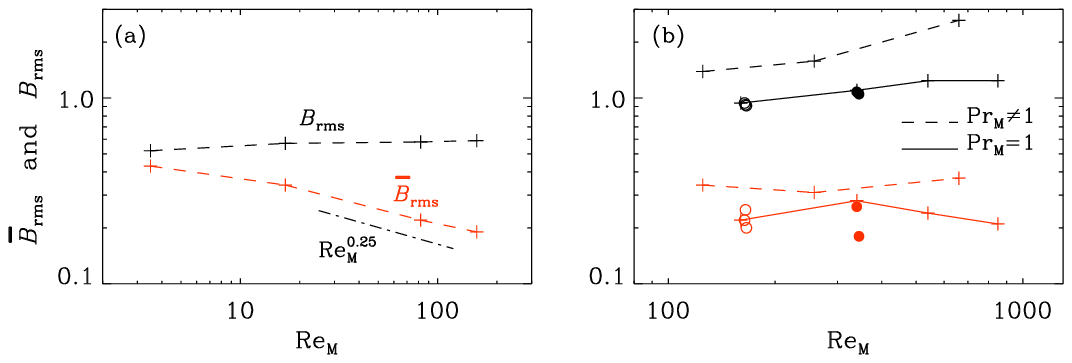}
\caption{
Dependence of $\Brms$ (black) and $\meanBrms$ (red) on $\Rm$
(a) without shear and (b) with shear.
Dashed lines indicate that $\Rm$ is varied by changing $\Pm$ [Runs~A--D
in (a) and Runs~E--G in (b)], while solid lines indicate that $\Rey$
has been changed [Runs~H--K in (b)].
In (b), the open and closed circles are for Runs~L--N and O+P,
respectively.
\label{fig:pb_vs_rm}}
\end{figure*}

\subsection{Superequipartition with shear} \label{sec:super}

The reason why our runs with shear show strong small-scale magnetic
helicity fluxes is probably connected with the fact that in those runs,
the magnetic field reaches superequipartition strengths.
This is seen in \Tab{Tsummary2}, where $\Brms/\Beq>1$, and in
\Fig{fig:pBm_two_E256a_5em5_q05b_rep}, where we plot for Run~E the
temporal evolution of $\bra{\BB^2}$, $\Beq^2=\mu_0\rho_0\,\urms^2$,
$\bra{B_y^2}$, and $\bra{B_x^2}$.
We see that the total (small-scale and large-scale) magnetic field
reaches superequipartition field strengths at $t\approx300/\cs k_1$,
which is clearly before the large-scale magnetic field saturates at
$t\approx1000/\cs k_1$, which is when $\bra{B_y^2}$ has reached a
statistically steady state.
Interestingly, the large-scale magnetic field displays an approximately
exponential growth at a rate $10^{-2}\,\urms\kf$ for the squared
mean-field strength.
The growth rate of the mean field is then half that value, which is
comparable to the growth rates found in \Tab{TMFS}, where
$\lambda/\cs k_1\approx0.005$, corresponding to
$\lambda/\urms\kf\approx0.01$.

The slow exponential growth of the mean field may hint at a new type
of instability that is responsible for the emergence of the large-scale
magnetic field in a regime in which the total magnetic field reaches
superequipartition strengths.
Analogous evidence for exponential growth of a secondary mean-field
instability has previously been seen in other circumstances; see
\cite{Bra+11} for an example in the context of density-stratified
turbulence in which large-scale structures were found to form.

In the shear-induced superequipartition regime, the rms values of the
resulting large-scale magnetic field are found to be independent of $\Rm$.
In the absence of shear, $\meanBrms$ clearly declined with $\Rm$, albeit
only like $\Rm^{0.3}$; see \Fig{fig:pb_vs_rm}(a).
With shear, however, $\meanBrms$ is nearly independent of $\Rm$; see
\Fig{fig:pb_vs_rm}(b).
The fact that resistive contributions through the integrated
$2\eta\mu_0\overline{\jj\cdot\bb}$ term become increasingly important can
be regarded as a consequence of the shear-induced hemispheric magnetic
helicity fluxes that become extremely efficient at exchanging small-scale
magnetic helicity between hemispheres.
We can therefore say that they overcome catastrophic quenching so as to
guarantee an $\Rm$-independent large-scale magnetic field.

\begin{figure}\begin{center}
\plotone{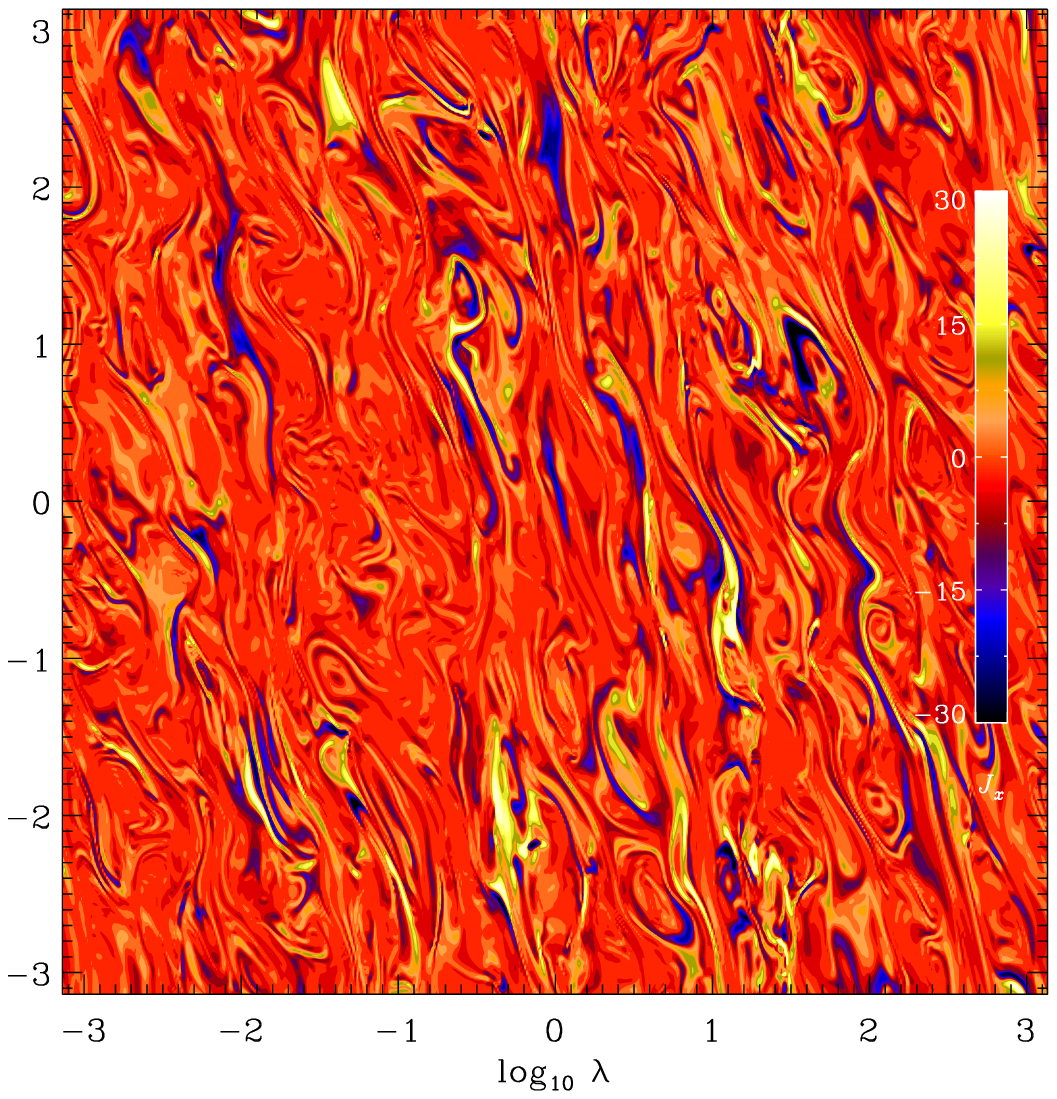}
\end{center}\caption{
Slice of $J_x(x,y,z_\ast)$ for Run~G at $k_1 z_\ast=1$, showing
a systematic tilt from the upper left to the lower right, with
all structures being well resolved.
}\label{jx2_slice_z1_E1024a_1em5_q05a}\end{figure}

The strong contribution of the small-scale current helicity term might
raise concerns whether the simulation is sufficiently well resolved.
To check this, we plot in \Fig{jx2_slice_z1_E1024a_1em5_q05a} an $xy$
slice of $J_x$ through $k_1 z=1$, but no signs of ringing, i.e.,
no oscillations on the grid scale are seen.
Instead, it shows the typical inclined patterns associated with the
shear flow, $\meanU_y(x)=-q\Omega x$.

\begin{figure}[t!]
\plotone{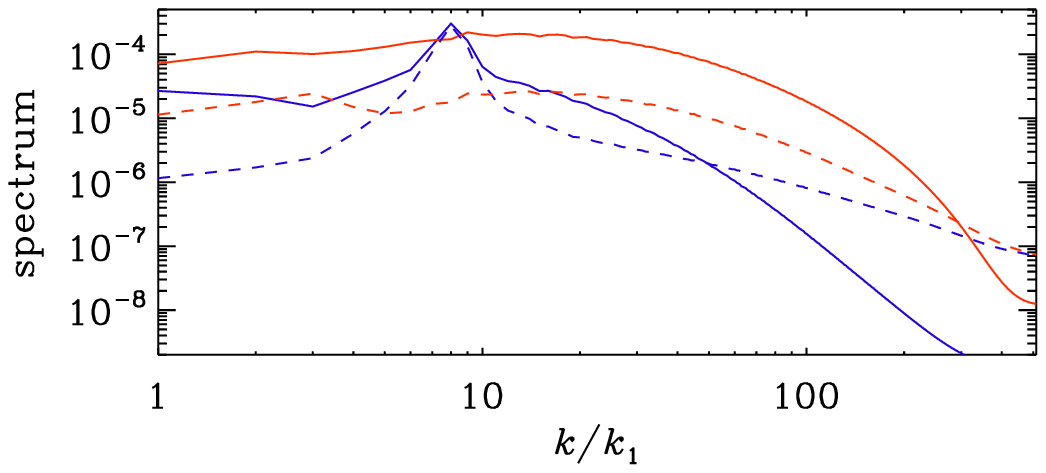}
\caption{
Three-dimensional kinetic (blue) and magnetic (red) energy spectra for
Runs~G (solid lines) and K (dashed lines).
\label{fig:pspec}}
\end{figure}

In \Fig{fig:pspec}, we present kinetic and magnetic energy spectra for
Runs~G and K.
Both runs have shear but different values of $\Pm$.
The spectra are similar, except that the magnetic and velocity spectra
for Run~K still have more energy at the largest wavenumber of the mesh,
i.e., at the Nyquist wavenumber $k_1 N/2$.

\begin{figure}[t!]
\plotone{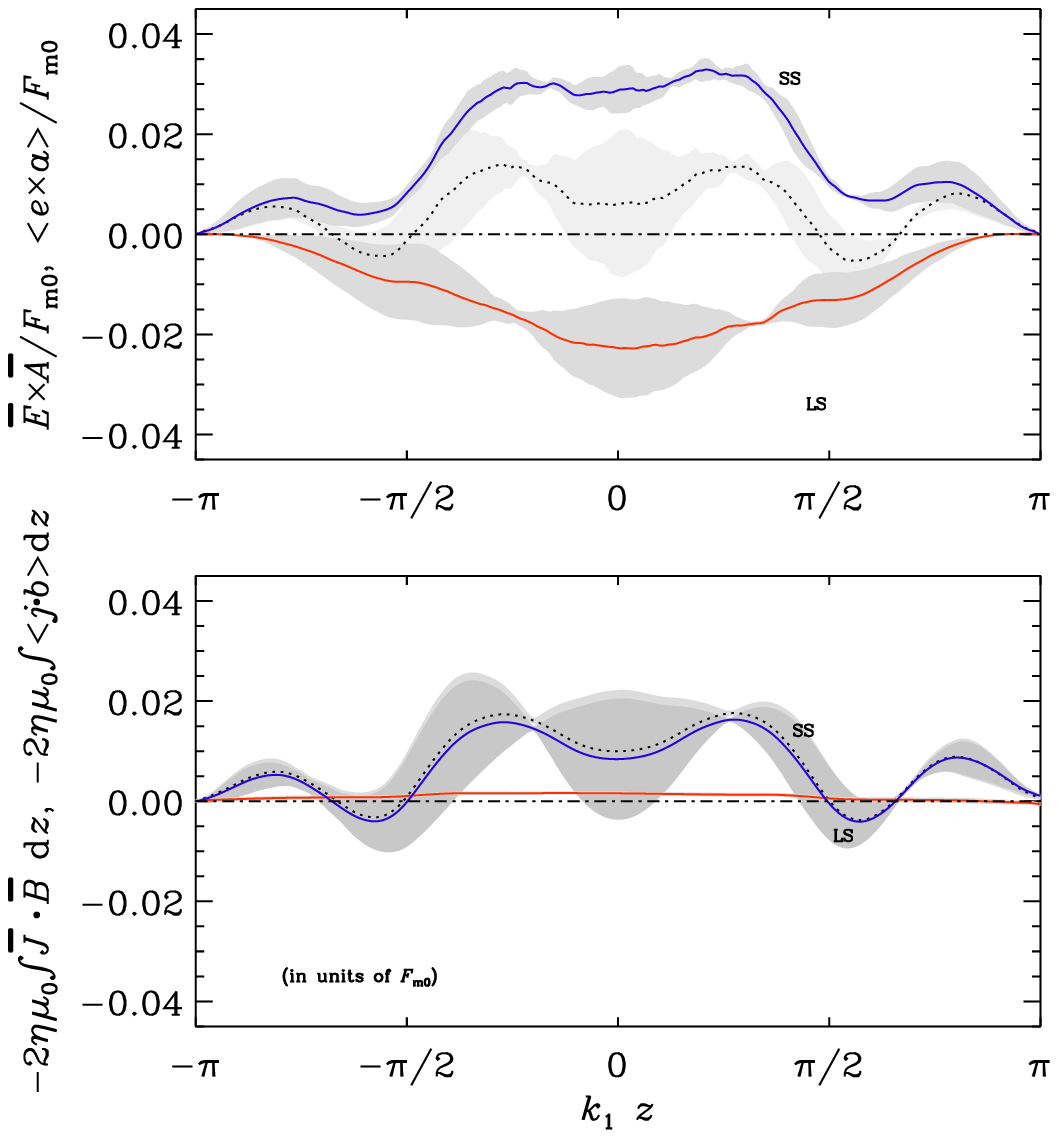}
\caption{
Magnetic helicity fluxes for Run~Q with a top-hat velocity profile.
As in \Fig{fig:pphelflux_E256a_5em5b},
the blue (red) lines denote the small-scale (large-scale) contributions
and the black dotted lines denote their sum.
The black dashed--dotted line is the zero line.
Note the sharp flanks of $\overline{\ee\times\aaaa}$ at $z=\pm\pi/2$.
\label{fig:pphelflux_2panels2_E256a_5em5_q0p5_step_new}}
\end{figure}

It is important to emphasize that only the total magnetic field and
not the large-scale field reaches superequipartition field strengths.
Such strong magnetic fields appear to be crucial for achieving the new
type of magnetic helicity fluxes explored in this paper.
Such fluxes are the result of anisotropies in the turbulence, and many
forms have been explored in earlier papers \citep{Vishniac+Cho01,
Subramanian+Brandenburg04, Subramanian+Brandenburg06, HB11,
Shapovalov+Vishniac11, DSGB13, Vishniac+Shapovalov14,
Ebrahimi+Bhattacharjee14, Zhou+Blackman17, KR22,
Gopalakrishnan+Subramanian23}.
However, more analytical work is needed to make meaningful statements
about the physical nature of the fluxes discussed in the present work.

\subsection{Sensitivity to a steeper profile and larger shear} \label{sec:SteeperProfile}

As alluded to in \Sec{sec:equations}, we adopted a sinusoidal modulation
for the turbulent intensity.
To get an idea about the sensitivity of our results upon this choice,
we now consider the top-hat profile defined in \Sec{sec:equations}.
The result for Run~Q is shown in
\Fig{fig:pphelflux_2panels2_E256a_5em5_q0p5_step_new}.
The differences to Run~E in \Fig{fig:pphelflux_E256a_5em5_q05b} are
relatively minor, except that the flanks of $\overline{\ee\times\aaaa}$
at $z=\pm\pi/2$ show a sharper onset at $z=\pm\pi/2$.
Also, the profile of the integrated $2\eta\mu_0\overline{\jj\cdot\bb}$ term
shows marked humps at these positions.

Another particular choice was that of the value of $q$, which corresponds
to the local double-logarithmic shear derivative in other astrophysical
contexts.
In astrophysical bodies such as stars, the value of $q$ can have either
sign in different regions.
Values in the range from $\pm0.1$ to $\pm1$ are not uncommon.
In galactic disks with a constant linear velocity, we have $q=1$,
whereas in Keplerian disks, we have $q=3/2$.
As discussed in \Sec{sec:equations}, with our choice of angular velocity
and domain size, our compressible simulations yield transonic velocities
for $q>0.5$.
To assess the sensitivity of our results upon the choice $q=0.5$, we
now consider a simulation similar to Run~E, but with $q=3/2$ and
a larger sound speed of $\cs=3$.
The result is listed in \Tab{Tsummary2} as Run~R.
The values of $\Rey$ and $\Rm$ are larger than for Run~E, because we
left the viscosity and magnetic diffusivity unchanged.
The kinetic helicity parameter is now nearly vanishing, but the current
helicity parameter remains unchanged compared to that for Run~E.
All the other parameters describing the mean magnetic field and magnetic
helicity fluxes are also nearly unchanged.
We therefore conclude that our particular choice of the value of $q$
had no decisive effect on the results.

\section{Conclusions}
\label{sec:Concl}

In this paper, we have considered a volume containing a modulated level
of turbulence, which drops to zero at the vertical boundaries, so that
 the internal transport coefficients of the medium
go to zero smoothly at those boundaries.
Consequently, the results of the simulations should be insensitive to
the vertical boundary conditions.
The properties of the medium, including rotation and shear, are otherwise
uniform within the volume.

The existence of a systematic flux of small-scale
helicity, as seen in \Fig{fig:pphelflux_E256a_5em5_q05b}, is consistent
with a simple dimensional estimate.
A magnetic helicity flux arising from turbulence should scale with
$\overline{\uu^2}\overline{\BB^2}\tau$, where $\tau$ is a correlation
time \citep{KR22}.
However, the magnetic helicity flux is a pseudo-vector, with a direction
that requires a scaling with either the rotation or the shear (or more
particularly with the local vorticity), which implies an extra factor
of $\Omega\tau$ or $S\tau$.
For a uniform driving scale of the turbulence, and a uniform shear and
rotation, we expect a magnetic helicity flux proportional to
$\overline{\BB^2}$, which is consistent with the evidence in case E.
Our results suggest that this flux is significantly more sensitive to shear than to rotation.

We have shown that in a large-scale dynamo, the magnetic helicity
fluxes between large and small scales can even become overcompensated by
those between the two hemispheres when microphysical resistive effects
are small.
In the absence of shear and at small magnetic Reynolds numbers (Run~A,
$\Rm=3.5$), these fluxes are comparable to the reference flux defined
in \Eq{RefFlux}.
For Run~D with $\Rm=160$, the magnetic helicity fluxes are about 30\%
of the reference flux.
However, while the magnetic helicity flux at large scales
is large, that at small scales is virtually absent; see
\Figs{fig:pphelflux_E256a_5em5b}{fig:pphelflux_TE256d}.
In the presence of shear and for similar values of $\Rm$ (Run~E),
the flux drops to 1--5\% of the reference flux, but now the fluxes are
approximately the same at large and small scales.

The correspondence between the magnetic helicity flux between hemispheres
and between scales is not a coincidence.
The small-scale flux between hemispheres depends on the total magnetic
field strength, although the small- and large-scale magnetic fields
contribute to the flux with different coefficients.
The flow of magnetic helicity between scales is proportional to the
square of the large-scale field.
Consequently, the saturation strength of the large-scale field may be
set by the magnetic helicity flux between hemispheres.

While the large fraction of small-scale magnetic helicity fluxes in the
presence of shear is indicative of alleviating catastrophic quenching,
we do not find that the resistive term becomes unimportant at large
magnetic Reynolds numbers.
This does not a priori mean that such dynamos are not viable in the
large-$\Rm$ regime.
The fact that $2\int\meanEMF\cdot\meanBB\,\dd z\to0$ in the large magnetic
Reynolds number limit was thought to reflect the basic catastrophic
quenching problem of large-scale dynamos with helicity.
However, while the hemispheric small-scale magnetic helicity flux
stays constant as $\Rm$ increases, it is not being used to balance the
integrated $2\meanEMF\cdot\meanBB$ term, but it either drives or is
driven by the integrated $2\eta\mu_0\overline{\jj\cdot\bb}$ term.

Our models have demonstrated that interesting flux dynamics can occur
entirely without boundaries.
The magnetic helicity fluxes occur within the volume due to gradients
of the turbulent intensity and turbulent kinetic and magnetic helicities,
as was also found by \cite{KR22}.

Our work has also shown that the dynamos in the present setups are of
$\alpha^2$ or $\alpha\Omega$ type, i.e., they are driven by an
$\alpha$ effect and supported by shear, if shear is present.
The ratio of the local value of $\alpha$ to the product of the local
turbulent magnetic diffusivity and the lowest wavenumber of the
domain, i.e., the $z$-dependent dynamo number, which is found to be
an approximately linear function with a coefficient of 2.8 and 2.2 for
Runs~A and D, respectively.
This suggests that the dynamo number decreases with increasing conductivity.
For Run~E with shear, the ratio is 1.4.
Turbulent pumping points in the direction away from the midplane.
There is also a R\"adler effect with the theoretically expected sign,
i.e., $\meanEMF$ has a contribution proportional to $\OO\times\meanJJ$
with a positive coefficient.
In the presence of shear, the effect becomes anisotropic and the
component that is relevant for the shear--current effect, namely the
$\eta_{yx}$ component for shear of the present form $S=\dd\meanU_y/\dd x$,
is essentially zero, which is consistent with earlier findings
\citep{BRRK08, Kapy+22}.

Our results have applications to stellar and galactic dynamos, where
gradients of kinetic and magnetic helicity fluxes are expected to occur
through the equator.
Even the boundary between convecting and nonconvecting regions both
in late-type stars and in massive stars is an example where magnetic
helicity fluxes can be expected to encounter dynamical boundaries of the type idealized here.

\begin{acknowledgments}
We thank the reviewer for a positive and constructive review.
We acknowledge inspiring discussions with the participants of
the program on ``Turbulence in Astrophysical Environments'' at the Kavli
Institute for Theoretical Physics in Santa Barbara.
This research was supported in part by the
Swedish Research Council (Vetenskapsr{\aa}det) under Grant No.\ 2019-04234,
the National Science Foundation under Grant Nos.\ NSF PHY-2309135 and AST-2307698,
and the NASA ATP Award 80NSSC22K0825.
We acknowledge the allocation of computing resources provided by the
Swedish National Allocations Committee at the Center for
Parallel Computers at the Royal Institute of Technology in Stockholm. ETV's participation in the KITP workshop and in this work specifically has been supported in part by funds provided by the American Astronomical Society.

\vspace{2mm}\noindent
{\em Software and Data Availability.}
The source code used for the simulations of this study,
the {\sc Pencil Code} \citep{PC}, is freely available on
\url{https://github.com/pencil-code/}.
The simulation setups and corresponding input
and reduced out data are freely available on
\dataset[http://doi.org/10.5281/zenodo.14968754]{http://doi.org/10.5281/zenodo.14974165}.

\end{acknowledgments}

\bibliography{ref}{}

\begin{thebibliography}{}
\expandafter\ifx\csname natexlab\endcsname\relax\def\natexlab#1{#1}\fi
\providecommand{\url}[1]{\href{#1}{#1}}
\providecommand{\dodoi}[1]{doi:~\href{http://doi.org/#1}{\nolinkurl{#1}}}
\providecommand{\doeprint}[1]{\href{http://ascl.net/#1}{\nolinkurl{http://ascl.net/#1}}}
\providecommand{\doarXiv}[1]{\href{https://arxiv.org/abs/#1}{\nolinkurl{https://arxiv.org/abs/#1}}}

\bibitem[{{Armitage}(2011)}]{Armitage11}
{Armitage}, P.~J. 2011, \araa, 49, 195,
  \dodoi{10.1146/annurev-astro-081710-102521}

\bibitem[{{Balbus} \& {Hawley}(1998)}]{BH98}
{Balbus}, S.~A., \& {Hawley}, J.~F. 1998, Rev. Mod. Phys., 70, 1,
  \dodoi{10.1103/RevModPhys.70.1}

\bibitem[{{Beck} {et~al.}(1996){Beck}, {Brandenburg}, {Moss}, {Shukurov}, \&
  {Sokoloff}}]{Beck+96}
{Beck}, R., {Brandenburg}, A., {Moss}, D., {Shukurov}, A., \& {Sokoloff}, D.
  1996, \araa, 34, 155, \dodoi{10.1146/annurev.astro.34.1.155}

\bibitem[{{Bermudez} \& {Alexakis}(2022)}]{Bermudez+Alexakis22}
{Bermudez}, G., \& {Alexakis}, A. 2022, \prl, 129, 195101,
  \dodoi{10.1103/PhysRevLett.129.195101}

\bibitem[{{Blackman} \& {Brandenburg}(2002)}]{BB02}
{Blackman}, E.~G., \& {Brandenburg}, A. 2002, \apj, 579, 359,
  \dodoi{10.1086/342705}

\bibitem[{{Blackman} \& {Field}(2000)}]{BF00}
{Blackman}, E.~G., \& {Field}, G.~B. 2000, \apj, 534, 984,
  \dodoi{10.1086/308767}

\bibitem[{{Brandenburg}(2001)}]{Bra01}
{Brandenburg}, A. 2001, \apj, 550, 824, \dodoi{10.1086/319783}

\bibitem[{{Brandenburg}(2005)}]{Bra05QPO}
---. 2005, Astron. Nachr., 326, 787, \dodoi{10.1002/asna.200510414}

\bibitem[{{Brandenburg}(2017)}]{Bra17}
---. 2017, \aap, 598, A117, \dodoi{10.1051/0004-6361/201630033}

\bibitem[{{Brandenburg}(2018{\natexlab{a}})}]{Bra18}
---. 2018{\natexlab{a}}, Astron. Nachr., 339, 631,
  \dodoi{10.1002/asna.201913604}

\bibitem[{{Brandenburg}(2018{\natexlab{b}})}]{Bran18}
---. 2018{\natexlab{b}}, J. Plasma Phys., 84, 735840404,
  \dodoi{10.1017/S0022377818000806}

\bibitem[{{Brandenburg} \& {Campbell}(1997)}]{BC97}
{Brandenburg}, A., \& {Campbell}, C. 1997, in Accretion Disks - New Aspects,
  ed. E.~{Meyer-Hofmeister} \& H.~{Spruit}, Vol. 487, 109,
  \dodoi{10.1007/BFb0105826}

\bibitem[{{Brandenburg} \& {Chatterjee}(2018)}]{BC18}
{Brandenburg}, A., \& {Chatterjee}, P. 2018, Astron. Nachr., 339, 118,
  \dodoi{10.1002/asna.201813472}

\bibitem[{{Brandenburg} {et~al.}(2013){Brandenburg}, {Gressel},
  {K{\"a}pyl{\"a}}, {Kleeorin}, {Mantere}, \& {Rogachevskii}}]{Bran+13}
{Brandenburg}, A., {Gressel}, O., {K{\"a}pyl{\"a}}, P.~J., {et~al.} 2013, \apj,
  762, 127, \dodoi{10.1088/0004-637X/762/2/127}

\bibitem[{{Brandenburg} {et~al.}(2011){Brandenburg}, {Kemel}, {Kleeorin},
  {Mitra}, \& {Rogachevskii}}]{Bra+11}
{Brandenburg}, A., {Kemel}, K., {Kleeorin}, N., {Mitra}, D., \& {Rogachevskii},
  I. 2011, \apjl, 740, L50, \dodoi{10.1088/2041-8205/740/2/L50}

\bibitem[{{Brandenburg} {et~al.}(1995){Brandenburg}, {Nordlund}, {Stein}, \&
  {Torkelsson}}]{BNST95}
{Brandenburg}, A., {Nordlund}, A., {Stein}, R.~F., \& {Torkelsson}, U. 1995,
  \apj, 446, 741, \dodoi{10.1086/175831}

\bibitem[{{Brandenburg} \& {Ntormousi}(2023)}]{BN23}
{Brandenburg}, A., \& {Ntormousi}, E. 2023, \araa, 61, 561,
  \dodoi{10.1146/annurev-astro-071221-052807}

\bibitem[{{Brandenburg} {et~al.}(2008{\natexlab{a}}){Brandenburg},
  {R{\"a}dler}, {Rheinhardt}, \& {K{\"a}pyl{\"a}}}]{BRRK08}
{Brandenburg}, A., {R{\"a}dler}, K.~H., {Rheinhardt}, M., \& {K{\"a}pyl{\"a}},
  P.~J. 2008{\natexlab{a}}, \apj, 676, 740, \dodoi{10.1086/527373}

\bibitem[{{Brandenburg} {et~al.}(2008{\natexlab{b}}){Brandenburg},
  {R{\"a}dler}, {Rheinhardt}, \& {Subramanian}}]{BRRS08}
{Brandenburg}, A., {R{\"a}dler}, K.-H., {Rheinhardt}, M., \& {Subramanian}, K.
  2008{\natexlab{b}}, \apjl, 687, L49, \dodoi{10.1086/593146}

\bibitem[{{Brandenburg} {et~al.}(2008{\natexlab{c}}){Brandenburg},
  {R{\"a}dler}, \& {Schrinner}}]{BRS08}
{Brandenburg}, A., {R{\"a}dler}, K.~H., \& {Schrinner}, M. 2008{\natexlab{c}},
  \aap, 482, 739, \dodoi{10.1051/0004-6361:200809365}

\bibitem[{{Brandenburg} \& {Scannapieco}(2020)}]{BS20}
{Brandenburg}, A., \& {Scannapieco}, E. 2020, \apj, 889, 55,
  \dodoi{10.3847/1538-4357/ab5e7f}

\bibitem[{{Candelaresi} {et~al.}(2011){Candelaresi}, {Hubbard}, {Brandenburg},
  \& {Mitra}}]{Can+11}
{Candelaresi}, S., {Hubbard}, A., {Brandenburg}, A., \& {Mitra}, D. 2011, Phys.
  Plasmas, 18, 012903, \dodoi{10.1063/1.3533656}

\bibitem[{{Cattaneo} \& {Hughes}(1996)}]{CH96}
{Cattaneo}, F., \& {Hughes}, D.~W. 1996, \pre, 54, R4532,
  \dodoi{10.1103/PhysRevE.54.R4532}

\bibitem[{{Charbonneau}(2014)}]{Charbonneau14}
{Charbonneau}, P. 2014, \araa, 52, 251,
  \dodoi{10.1146/annurev-astro-081913-040012}

\bibitem[{{Cole} {et~al.}(2016){Cole}, {Brandenburg}, {K{\"a}pyl{\"a}}, \&
  {K{\"a}pyl{\"a}}}]{Cole+16}
{Cole}, E., {Brandenburg}, A., {K{\"a}pyl{\"a}}, P.~J., \& {K{\"a}pyl{\"a}},
  M.~J. 2016, \aap, 593, A134, \dodoi{10.1051/0004-6361/201628165}

\bibitem[{{Davis} \& {Tchekhovskoy}(2020)}]{DT20}
{Davis}, S.~W., \& {Tchekhovskoy}, A. 2020, \araa, 58, 407,
  \dodoi{10.1146/annurev-astro-081817-051905}

\bibitem[{{Del Sordo} {et~al.}(2013){Del Sordo}, {Guerrero}, \&
  {Brandenburg}}]{DSGB13}
{Del Sordo}, F., {Guerrero}, G., \& {Brandenburg}, A. 2013, \mnras, 429, 1686,
  \dodoi{10.1093/mnras/sts398}

\bibitem[{{Ebrahimi} \& {Bhattacharjee}(2014)}]{Ebrahimi+Bhattacharjee14}
{Ebrahimi}, F., \& {Bhattacharjee}, A. 2014, \prl, 112, 125003,
  \dodoi{10.1103/PhysRevLett.112.125003}

\bibitem[{{Field} \& {Blackman}(2002)}]{FB02}
{Field}, G.~B., \& {Blackman}, E.~G. 2002, \apj, 572, 685,
  \dodoi{10.1086/340233}

\bibitem[{{Gopalakrishnan} \&
  {Subramanian}(2023)}]{Gopalakrishnan+Subramanian23}
{Gopalakrishnan}, K., \& {Subramanian}, K. 2023, \apj, 943, 66,
  \dodoi{10.3847/1538-4357/aca808}

\bibitem[{{Gressel} {et~al.}(2008){Gressel}, {Ziegler}, {Elstner}, \&
  {R{\"u}diger}}]{Gressel+08}
{Gressel}, O., {Ziegler}, U., {Elstner}, D., \& {R{\"u}diger}, G. 2008, Astron.
  Nachr., 329, 619, \dodoi{10.1002/asna.200811005}

\bibitem[{{Gruzinov} \& {Diamond}(1996)}]{GD96}
{Gruzinov}, A.~V., \& {Diamond}, P.~H. 1996, Phys. Plasmas, 3, 1853,
  \dodoi{10.1063/1.871981}

\bibitem[{{Haugen} {et~al.}(2004){Haugen}, {Brandenburg}, \& {Dobler}}]{HBD04}
{Haugen}, N.~E., {Brandenburg}, A., \& {Dobler}, W. 2004, \pre, 70, 016308,
  \dodoi{10.1103/PhysRevE.70.016308}

\bibitem[{{Hubbard} \& {Brandenburg}(2009)}]{HB09}
{Hubbard}, A., \& {Brandenburg}, A. 2009, \apj, 706, 712,
  \dodoi{10.1088/0004-637X/706/1/712}

\bibitem[{{Hubbard} \& {Brandenburg}(2010)}]{HB10}
---. 2010, GApFD, 104, 577, \dodoi{10.1080/03091929.2010.506438}

\bibitem[{{Hubbard} \& {Brandenburg}(2011)}]{HB11}
---. 2011, \apj, 727, 11, \dodoi{10.1088/0004-637X/727/1/11}

\bibitem[{{Hubbard} \& {Brandenburg}(2012)}]{HB12}
---. 2012, \apj, 748, 51, \dodoi{10.1088/0004-637X/748/1/51}

\bibitem[{{Jabbari} {et~al.}(2014){Jabbari}, {Brandenburg}, {Losada},
  {Kleeorin}, \& {Rogachevskii}}]{Jab+14}
{Jabbari}, S., {Brandenburg}, A., {Losada}, I.~R., {Kleeorin}, N., \&
  {Rogachevskii}, I. 2014, \aap, 568, A112, \dodoi{10.1051/0004-6361/201423499}

\bibitem[{{Ji}(1999)}]{Ji99}
{Ji}, H. 1999, \prl, 83, 3198, \dodoi{10.1103/PhysRevLett.83.3198}

\bibitem[{{Jiang} {et~al.}(2014){Jiang}, {Stone}, \& {Davis}}]{Jiang+14}
{Jiang}, Y.-F., {Stone}, J.~M., \& {Davis}, S.~W. 2014, \apj, 796, 106,
  \dodoi{10.1088/0004-637X/796/2/106}

\bibitem[{{K{\"a}pyl{\"a}} {et~al.}(2022){K{\"a}pyl{\"a}}, {Rheinhardt}, \&
  {Brandenburg}}]{Kapy+22}
{K{\"a}pyl{\"a}}, M.~J., {Rheinhardt}, M., \& {Brandenburg}, A. 2022, \apj,
  932, 8, \dodoi{10.3847/1538-4357/ac5b78}

\bibitem[{{K{\"a}pyl{\"a}} {et~al.}(2020){K{\"a}pyl{\"a}}, {Vizoso},
  {Rheinhardt}, {Brandenburg}, \& {Singh}}]{Kapy+20}
{K{\"a}pyl{\"a}}, M.~J., {Vizoso}, J.~{\'A}., {Rheinhardt}, M., {Brandenburg},
  A., \& {Singh}, N.~K. 2020, \apj, 905, 179, \dodoi{10.3847/1538-4357/abc1e8}

\bibitem[{{Karak} {et~al.}(2014){Karak}, {Rheinhardt}, {Brandenburg},
  {K{\"a}pyl{\"a}}, \& {K{\"a}pyl{\"a}}}]{Karak+14}
{Karak}, B.~B., {Rheinhardt}, M., {Brandenburg}, A., {K{\"a}pyl{\"a}}, P.~J.,
  \& {K{\"a}pyl{\"a}}, M.~J. 2014, \apj, 795, 16,
  \dodoi{10.1088/0004-637X/795/1/16}

\bibitem[{{Kleeorin} {et~al.}(2000){Kleeorin}, {Moss}, {Rogachevskii}, \&
  {Sokoloff}}]{Kleeorin+00}
{Kleeorin}, N., {Moss}, D., {Rogachevskii}, I., \& {Sokoloff}, D. 2000, \aap,
  361, L5, \dodoi{10.48550/arXiv.astro-ph/0205266}

\bibitem[{{Kleeorin} \& {Rogachevskii}(2022)}]{KR22}
{Kleeorin}, N., \& {Rogachevskii}, I. 2022, \mnras, 515, 5437,
  \dodoi{10.1093/mnras/stac2141}

\bibitem[{{Krause} \& {R{\"a}dler}(1980)}]{KR80}
{Krause}, F., \& {R{\"a}dler}, K.-H. 1980, {Mean-Field Magnetohydrodynamics and
  Dynamo Theory} (Oxford: Pergamon Press)

\bibitem[{{Moffatt}(1978)}]{Mof78}
{Moffatt}, H.~K. 1978, {Magnetic Field Generation in Electrically Conducting
  Fluids} (Cambridge: Cambridge University Press)

\bibitem[{{Ossendrijver} {et~al.}(2002){Ossendrijver}, {Stix}, {Brandenburg},
  \& {R{\"u}diger}}]{Ossendrijver+02}
{Ossendrijver}, M., {Stix}, M., {Brandenburg}, A., \& {R{\"u}diger}, G. 2002,
  \aap, 394, 735, \dodoi{10.1051/0004-6361:20021224}

\bibitem[{{Parker}(1955)}]{Par55}
{Parker}, E.~N. 1955, \apj, 122, 293, \dodoi{10.1086/146087}

\bibitem[{{Parker}(1979)}]{Par79}
---. 1979, {Cosmical Magnetic Fields: Their Origin and Their Activity} (Oxford:
  Clarendon Press)

\bibitem[{{Pencil Code Collaboration} {et~al.}(2021){Pencil Code
  Collaboration}, {Brandenburg}, {Johansen}, {Bourdin}, {Dobler}, {Lyra},
  {Rheinhardt}, {Bingert}, {Haugen}, {Mee}, {Gent}, {Babkovskaia}, {Yang},
  {Heinemann}, {Dintrans}, {Mitra}, {Candelaresi}, {Warnecke},
  {K{\"a}pyl{\"a}}, {Schreiber}, {Chatterjee}, {K{\"a}pyl{\"a}}, {Li},
  {Kr{\"u}ger}, {Aarnes}, {Sarson}, {Oishi}, {Schober}, {Plasson}, {Sandin},
  {Karchniwy}, {Rodrigues}, {Hubbard}, {Guerrero}, {Snodin}, {Losada},
  {Pekkil{\"a}}, \& {Qian}}]{PC}
{Pencil Code Collaboration}, {Brandenburg}, A., {Johansen}, A., {et~al.} 2021,
  JOSS, 6, 2807, \dodoi{10.21105/joss.02807}

\bibitem[{{Pipin}(2023)}]{Pipin23}
{Pipin}, V.~V. 2023, \mnras, 522, 2919, \dodoi{10.1093/mnras/stad1150}

\bibitem[{{Pouquet} {et~al.}(1976){Pouquet}, {Frisch}, \& {Leorat}}]{PFL76}
{Pouquet}, A., {Frisch}, U., \& {Leorat}, J. 1976, JFM, 77, 321,
  \dodoi{10.1017/S0022112076002140}

\bibitem[{{R{\"a}dler}(1969)}]{Radler69}
{R{\"a}dler}, K.~H. 1969, Monatsber. Deutsch. Akad Wissensch. Berlin, 11, 194

\bibitem[{{Rheinhardt} \& {Brandenburg}(2010)}]{RB10}
{Rheinhardt}, M., \& {Brandenburg}, A. 2010, \aap, 520, A28,
  \dodoi{10.1051/0004-6361/201014700}

\bibitem[{{Rheinhardt} \& {Brandenburg}(2012)}]{RB12}
---. 2012, Astron. Nachr., 333, 71, \dodoi{10.1002/asna.201111625}

\bibitem[{{Rincon}(2021)}]{Rin21}
{Rincon}, F. 2021, PhRvF, 6, L121701, \dodoi{10.1103/PhysRevFluids.6.L121701}

\bibitem[{{Rogachevskii} \& {Kleeorin}(2003)}]{RK03}
{Rogachevskii}, I., \& {Kleeorin}, N. 2003, \pre, 68, 036301,
  \dodoi{10.1103/PhysRevE.68.036301}

\bibitem[{{Rogachevskii} \& {Kleeorin}(2004)}]{RK04}
---. 2004, \pre, 70, 046310, \dodoi{10.1103/PhysRevE.70.046310}

\bibitem[{{R{\"u}diger} \& {Brandenburg}(2014)}]{RB14}
{R{\"u}diger}, G., \& {Brandenburg}, A. 2014, \pre, 89, 033009,
  \dodoi{10.1103/PhysRevE.89.033009}

\bibitem[{{R\"udiger} \& {Kichatinov}(1993)}]{RK93}
{R\"udiger}, G., \& {Kichatinov}, L.~L. 1993, \aap, 269, 581

\bibitem[{{Schrinner} {et~al.}(2005){Schrinner}, {R{\"a}dler}, {Schmitt},
  {Rheinhardt}, \& {Christensen}}]{Sch05}
{Schrinner}, M., {R{\"a}dler}, K.~H., {Schmitt}, D., {Rheinhardt}, M., \&
  {Christensen}, U. 2005, Astron. Nachr., 326, 245,
  \dodoi{10.1002/asna.200410384}

\bibitem[{{Schrinner} {et~al.}(2007){Schrinner}, {R{\"a}dler}, {Schmitt},
  {Rheinhardt}, \& {Christensen}}]{Sch07}
{Schrinner}, M., {R{\"a}dler}, K.-H., {Schmitt}, D., {Rheinhardt}, M., \&
  {Christensen}, U.~R. 2007, GApFD, 101, 81, \dodoi{10.1080/03091920701345707}

\bibitem[{{Shapovalov} \& {Vishniac}(2011)}]{Shapovalov+Vishniac11}
{Shapovalov}, D.~S., \& {Vishniac}, E.~T. 2011, \apj, 738, 66,
  \dodoi{10.1088/0004-637X/738/1/66}

\bibitem[{{Shukurov} \& {Subramanian}(2022)}]{SS22}
{Shukurov}, A., \& {Subramanian}, K. 2022, Astrophysical Magnetic Fields: From
  Galaxies to the Early Universe (Cambridge: Cambridge University Press)

\bibitem[{{Squire} \& {Bhattacharjee}(2015)}]{Squire+Bhattacharjee15}
{Squire}, J., \& {Bhattacharjee}, A. 2015, \pre, 92, 053101,
  \dodoi{10.1103/PhysRevE.92.053101}

\bibitem[{{Steenbeck} {et~al.}(1966){Steenbeck}, {Krause}, \&
  {R{\"a}dler}}]{SKR66}
{Steenbeck}, M., {Krause}, F., \& {R{\"a}dler}, K.~H. 1966, Zeitschr.
  Naturforsch. A, 21, 369, \dodoi{10.1515/zna-1966-0401}

\bibitem[{{Stefani} \& {Gerbeth}(2003)}]{Stefani+Gerbeth03}
{Stefani}, F., \& {Gerbeth}, G. 2003, \pre, 67, 027302,
  \dodoi{10.1103/PhysRevE.67.027302}

\bibitem[{{Stepinski} \& {Levy}(1990)}]{SL90}
{Stepinski}, T.~F., \& {Levy}, E.~H. 1990, \apj, 362, 318,
  \dodoi{10.1086/169268}

\bibitem[{{Subramanian} \& {Brandenburg}(2004)}]{Subramanian+Brandenburg04}
{Subramanian}, K., \& {Brandenburg}, A. 2004, \prl, 93, 205001,
  \dodoi{10.1103/PhysRevLett.93.205001}

\bibitem[{{Subramanian} \& {Brandenburg}(2006)}]{Subramanian+Brandenburg06}
---. 2006, \apjl, 648, L71, \dodoi{10.1086/507828}

\bibitem[{{Vishniac} \& {Cho}(2001)}]{Vishniac+Cho01}
{Vishniac}, E.~T., \& {Cho}, J. 2001, \apj, 550, 752, \dodoi{10.1086/319817}

\bibitem[{{Vishniac} \& {Shapovalov}(2014)}]{Vishniac+Shapovalov14}
{Vishniac}, E.~T., \& {Shapovalov}, D. 2014, \apj, 780, 144,
  \dodoi{10.1088/0004-637X/780/2/144}

\bibitem[{{Warnecke} {et~al.}(2021){Warnecke}, {Rheinhardt}, {Viviani}, {Gent},
  {Tuomisto}, \& {K{\"a}pyl{\"a}}}]{Warnecke+21}
{Warnecke}, J., {Rheinhardt}, M., {Viviani}, M., {et~al.} 2021, \apjl, 919,
  L13, \dodoi{10.3847/2041-8213/ac1db5}

\bibitem[{{Zeldovich} {et~al.}(1983){Zeldovich}, {Ruzmaikin}, \&
  {Sokoloff}}]{ZRS83}
{Zeldovich}, {\relax Ya}.~B., {Ruzmaikin}, A.~A., \& {Sokoloff}, D.~D. 1983,
  {Magnetic Fields in Astrophysics} (New York: Gordon and Breach)

\bibitem[{{Zenati} \& {Vishniac}(2023)}]{ZV23}
{Zenati}, Y., \& {Vishniac}, E.~T. 2023, \apj, 948, 11,
  \dodoi{10.3847/1538-4357/acca1e}

\bibitem[{{Zhou} \& {Blackman}(2017)}]{Zhou+Blackman17}
{Zhou}, H., \& {Blackman}, E.~G. 2017, \mnras, 469, 1466,
  \dodoi{10.1093/mnras/stx914}

\bibitem[{{Zhou} \& {Blackman}(2024)}]{Zhou+Blackman24}
---. 2024, \pre, 109, 015206, \dodoi{10.1103/PhysRevE.109.015206}

\end{thebibliography}
\bibliographystyle{aasjournal}
\end{document}